\documentclass[review,onefignum,onetabnum]{siamonline250211}

\usepackage{lipsum}
\usepackage{amsfonts}
\usepackage{graphicx}
\usepackage{epstopdf}
\usepackage{algorithmic}
\ifpdf
  \DeclareGraphicsExtensions{.eps,.pdf,.png,.jpg}
\else
  \DeclareGraphicsExtensions{.eps}
\fi

\usepackage{enumitem}
\setlist[enumerate]{leftmargin=.5in}
\setlist[itemize]{leftmargin=.5in}

\newsiamremark{remark}{Remark}
\newsiamremark{defn}{Definition}
\newsiamremark{cor}{Corollary}
\newsiamremark{prop}{Proposition}
\newsiamremark{rem}{Remark}
\newsiamremark{assump}{Assumption}
\newsiamremark{hypothesis}{Hypothesis}
\crefname{hypothesis}{Hypothesis}{Hypotheses}
\newsiamthm{claim}{Claim}
\newsiamremark{fact}{Fact}
\crefname{fact}{Fact}{Facts}

\usepackage{mathtools}

\newcommand{\R}{\mathbb{R}}

\usepackage{mathrsfs}

\newcounter{hypA}
\newenvironment{hypA}{\refstepcounter{hypA}\begin{itemize}
  \item[({\bf A\arabic{hypA}})]}{\end{itemize}}
\newcounter{hypB}

\newcounter{hypD}

\newcounter{hypW}

\headers{Unbiased Approximations for McKean Vlasov SDEs}{E. Awadelkarim, N. K. Chada and A. Jasra}

\title{Unbiased Approximations for Stationary Distributions \\ of McKean-Vlasov SDEs\thanks{Submitted to the editors DATE.
\funding{AJ was supported by CUHK-SZ UDF01003537. NKC was supported by a City University of Hong Kong
Start-up Grant, project number: 7200809.}}}

\author{Elsiddig Awadelkarim\thanks{1Applied Mathematics and Computational Science Program, Computer, Electrical and Mathematical Sciences and
Engineering Division, King Abdullah University of Science and Technology, Thuwal, 23955-6900, KSA, (\email{elsiddigawadelkarim.elsiddig@kaust.edu.sa}.}
\and Neil K. Chada\thanks{ Department of Mathematics, City University of Hong Kong, HK-SAR.
  (\email{neilchada123@gmail.com)}.}
\and Ajay Jasra\thanks{ School of Data Science, The Chinese University of Hong Kong, Shenzhen, Shenzhen
  (\email{ajayjasra@cuhk.edu.cn)}.}}

\usepackage{amsopn}

\ifpdf
\hypersetup{
  pdftitle={An Example Article},
  pdfauthor={D. Doe, P. T. Frank, and J. E. Smith}
}
\fi

\nolinenumbers

\begin{document}

\maketitle

\begin{abstract}
We consider the development of unbiased estimators, to approximate the stationary distribution of Mckean-Vlasov stochastic differential
equations (MVSDEs). These are an important class of processes, which frequently appear in applications such as mathematical finance, biology and 
opinion dynamics.  Typically the stationary distribution is unknown and indeed
one cannot simulate such processes exactly.  As a result one commonly requires a time-discretization scheme which results in a discretization bias and a bias from not being able to simulate the associated stationary distribution.
To overcome this bias, we present a new unbiased estimator taking motivation from the literature on unbiased Monte Carlo.  We prove the unbiasedness of our estimator,  under certain assumptions. 
In order to prove this we require developing ergodicity results of various discrete time processes, through an appropriate discretization scheme,
towards the invariant measure.  Numerous numerical experiments are provided, on a range of MVSDEs, to demonstrate the effectiveness of our unbiased
estimator. Such examples include the Currie-Weiss model, a 3D neuroscience model and a parameter estimation problem.
\end{abstract}



\section{Introduction}
The focus of this article is on McKean-Vlasov stochastic differential equations (SDEs), which are SDEs
whose coefficients depend not only on the state of the process but also on its distribution.
In particular, we focus on the following McKean-Vlasov \cite{mck} stochastic differential equation (MVSDE), with a fixed initial condition $X_0=x_0\in\mathbb{R}^d$,
given as
\begin{equation}\label{eq:sde}
dX_t = a\left(X_t,\overline{\xi}_1(X_t,\mu_t)\right)dt + b\left(X_t,\overline{\xi}_2(X_t,\mu_t)\right)dW_t,
\end{equation}
where for $j\in\{1,2\}$,
$$
\overline{\xi}_j(X_t,\mu_t)  =  \int_{\mathbb{R}^d}\xi_j(X_t,x)\mu_t(dx),
$$
where $\{W_t\}_{t\geq 0}$ is a standard $d-$dimensional Brownian motion, for $d \geq 1$. Furthemore, $\xi_j:\mathbb{R}^{2d}\rightarrow\mathbb{R}$, $a:\mathbb{R}^d\times\mathbb{R}\rightarrow\mathbb{R}^d$ is the associated drift term, $b:\mathbb{R}^d\times\mathbb{R}\rightarrow\mathbb{R}^{d\times d}$ is the diffusion coefficient and finally $\mu_t$ is the law of the diffusion process $X_t$. 
In contrast to classical SDEs, the distribution of the MVSDE \eqref{eq:sde} solves a nonlinear Fokker-Planck equation, which is a partial differential equation.   McKean-Vlasov processes have proven to be very useful for  inference problems, in a wide range of useful applications. These applications include, but are not limited to,  stochastic filtering, financial mathematics, opinion dynamics and flocking processes.  Well-known examples of MVSDEs include the Kalman-Bucy filter, the stochastic Cucker-Smale flocking dynamics and the stochastic Hegselmann-Krause model  \cite{bucy,erban,garnier,motsch}.
Normally the solution of a MVSDE is approximated through an interacting particle system, where it is well-known that \textcolor{black}{the empirical law of the  particle system} converges to the law of the MVSDE, in the infinite particle limit.

We are interested in the simulation of the stationary distribution, or invariant measure, of the \textcolor{black}{MVSDE},  call it $\pi$,  which is assumed to exist, but is typically unknown.
\\
\textcolor{black}{There has been a recent surge in literature on providing methods to approximate $\pi$. This has been achieved using different methodologies. In a Bayesian framework, considerable work has been done on developing important sampling techniques. This originated in \cite{delmoral}, and has been extended in different setups \cite{ml_is_mv,mi_is_mv,reis}. Additional work has looked at convergence analysis, for more advanced methods such as HMC \cite{Bou}. For non-Bayesian techniques, we refer to the reader to the following works \cite{pages, du,HL23}, which also includes developing analysis, with particular assumptions stated. 
Finally, parameter estimation for MVSDEs has been a topic of interest as well \cite{chen,maama,sharrock}. There are many other interesting works in this field, related to the simulation of MVSDEs.
}

 However, despite these recent developments, there is still an issue related to these approximations methods, in that there remains a
time-discretization bias, through numerically solving \eqref{eq:sde}. Therefore, we motivate this work through the question of whether one can attain an unbiased scheme, to approximate invariant measures. In the context of Monte Carlo methods, unbiased estimation has been a recent hot topic largely due to important works by Rhee and Glynn. Specifically the authors produced an unbiased estimator related to SDEs through a randomized multilevel telescoping sum identity. We refer the reader to these various works \cite{glynn2,mcl,rhee}.

The notion of randomized multilevel Monte Monte Carlo (MLMC) methods is based upon a hierarchy of time-discretized SDEs becoming increasingly more precice in terms of discretization (i.e. an adaptive step-size).  Then given a probability distribution on the amount of time-discretization it is possible to obtain unbiased and sometimes finite variance,  finite expected cost estimators associated to ordinary SDEs \cite{ub_grad_new,ub_grad,ub_pf,rhee} and using non-randomized MLMC for some classes of MVSDE problems; \cite{siddig,ml_is_mv,mi_is_mv}. In terms of simulating invariant measures associated to regular SDEs several works have appeared including \cite{chada,langevin}.  The main idea of this article is to appropriately adapt and analyze methodology from 
\cite{siddig, ub_pf} in the context of producing truely unbiased estimation from the stationary distribution of a MVSDE model.

\subsection{Contributions}

The main contributions of this article are provided below:
\begin{itemize}
\item We develop a first approximation scheme for unbiased estimates of the stationary distribution of MVSDEs. This is based on the notion of randomization of MLMC, which utilizes two Euler-Maruyama discretizations of MVSDEs.
\item We prove a number of ergodicity results related to the Euler-Maruyama discretizations of the MVSDEs. In particular we demonstrate exponential ergodicity of both the discretized equations, 
and the unbiased estimator. To the best of the authors knowledge, these are the first set of results in the literature for the discretized setting.
\item We provide and present a main theorem which demonstrates that, our estimator is unbiased. This result relies heavily on the previous ergodicity results.
\item Numerical experiments are provided, to demonstrate the robustness of the proposed unbiased approximation scheme. We test this on a range of MVSDEs motivated through different applications. These include a parameter estimation problem, 
an Ornstein-Uhlenbeck process, the Currie-Weiss Model and a more challenging 3D neuron model.
\end{itemize}

The outline of this paper is as follows. In Section \ref{sec:method} we present our unbiased methodology.
Section \ref{sec:theory} houses our mathematical results with some discussion on how some simulation parameters of the method can be chosen.
Numerical experiments will be provided in Section \ref{sec:num}, to verify our theoretical findings. We will test our methodology on a range of MVSDEs which include a toy Ornstein-Uhlenbeck process, 3D neuron model (motivated from neuroscience) and the Currie-Weiss Model. We conclude with some final remarks in Section \ref{sec:conc}. Finally we defer the proofs of most of our results to the appendix.

\section{Method}

\label{sec:method}

\subsection{Overview of the Method and the Notation}
\textcolor{black}{In this section we will introduce and describe our unbiased methodology.  We will introduce our discretization, the notion of unbiased Monte Carlo, and relate it to MVSDEs. In particular,  we will use the convention that the time-discretized dynamics of the SDE \eqref{eq:sde} is iterated over unit time.  This is simply a convention and any  $\mathcal{O}(1)$ time could be used.  The reason for such iteration shall be explained later on.}
\textcolor{black}{Throughout the section, for ease of reading, we use $X_t$ to denote several processes, but we will always proceed the use of the notation by the definition of the process. The laws $\mu_t^l$ will denote the laws of the discretization we utilize below, while $\mu_t^{l,N}$ are empirical measures that we will generate in the algorithmsand constitutes a crucial step of the method. We use the notation $u_t$ with $t\in\mathbb{N}$ for the final processes that we will use to construct the estimators $\xi_l$.}

\subsection{Discretization for MVSDE}
Let us denote by $P_{\mu_{t-1},t}(x_{t-1},dx_t)$ the conditional law of $X_t$ (as given in \eqref{eq:sde}) given $\mathscr{F}_{t-1}$ (the natural filtration of the process), for $t\geq 1$; that is, the transition kernel over unit time. 
In most cases of practical interest, $\mu_t$ and the dynamics $P_{\mu_{t-1},t}(x_{t-1},dx_t)$ are difficult to work with.  For instance the transition kernel cannot be simulated in many problems.  We introduce a time-discretization over a regular grid of spacing $\Delta_l=2^{-l}$, $l\in\mathbb{N}_0$. 
We will use the Euler-Maruyama method associated to \eqref{eq:sde} and denote the law at any time $t\in\{0,\Delta_l,2\Delta_l,\dots\}$ as $\mu_t^l$. That is, we now consider the approximation for $k\in\mathbb{N}_0$:
\begin{align}
X_{k\Delta_l} & =  X_{(k-1)\Delta_l} + a\left(X_{(k-1)\Delta_l},\overline{\xi}_1(X_{(k-1)\Delta_l},\mu_{(k-1)\Delta_l}^l)\right) \textcolor{black}{\Delta_l} +\nonumber \\ & b\left(X_{(k-1)\Delta_l},\overline{\xi}_2(X_{(k-1)\Delta_l},\mu_{(k-1)\Delta_l}^l)\right)\left[W_{k\Delta_l} - W_{(k-1)\Delta_l}\right]\label{eq:euler}
\end{align}
where $X_0=x_0$ and $\mu_0^l=\delta_{\{x_0\}}$. \textcolor{black}{We note that the choice of $X_0=x_0$ is arbitrary, and be be both deterministic or random. As we will see later, certain conditions on the invariant measure will be independent of the choice of $x_0$. For the purposes of the numerical experiments these conditions are chosen from a random uniform distribution}.  Associated to \eqref{eq:euler}, we denote by $P_{\mu_{t-1}^l,t}^l(x_{t-1},dx_t)$ the conditional law of $X_t$, $t\in\mathbb{N}$,  given $\mathscr{F}_{t-1}$ for $t\geq 1$; that is, the transition kernel over unit time induced by \eqref{eq:euler}.  It should be remarked that in many cases, 
\eqref{eq:euler} cannot be simulated exactly as the expectations associated to $\mu_{(k-1)\Delta_l}^l$ cannot be computed even if one knows $\mu_{(k-1)\Delta_l}^l$, which is again unlikely.  We denote by $\pi^l$ stationary distribution of $P_{\mu_{t-1}^l,t}^l(x_{t-1},dx_t)$ which is assumed to exist.  More precisely,   we shall make some assumptions later on in the article which ensure that 
$$
\lim_{t\rightarrow\infty}\mathcal{W}_2(\mu_t^l,\pi^l) = 0,
$$
where $\mathcal{W}_2$ is the Wasserstein$-2$ distance (which will be defined later on).  We note that the mathematical details here are minimized to help readibility of this section of the article.

We now consider a method that will be used to provide a Monte Carlo based approximation of the law
$\mu_t^l$. The approach we present is a simple discretized method in Algorithm \ref{alg:basic_method} from \cite{basic_method}.   In Algorithm \ref{alg:basic_method}, the notation $\mathcal{N}_d(\kappa,\Sigma)$ denotes the $d-$dimensional Gaussian distribution with mean $\kappa$ and covariance matrix $\Sigma$. $I_d$ is the $d\times d$
identity matrix and $\stackrel{\textrm{ind}}{\sim}$ denotes independently distributed as.
Algorithm \ref{alg:basic_method} can be used to approximate expectations w.r.t.~$\mu_t^l$
and indeed on the grid in-between time $t-1$ and $t$.   Algorithm \ref{alg:basic_method} is given in a form that we require later in the article.  In our method, it will be critical that for two discretizations,  we will need to be able to sample from a dependent coupling of a pair of laws $\mu_t^l,\mu_t^{l-1}$. This is presented in Algorithm \ref{alg:basic_method_coup}.

\begin{algorithm}[h!]
\begin{enumerate}
\item{Input $l\in\mathbb{N}_0$ the level of discretization, $N\in\mathbb{N}$ the number of particles,  $t\in\{1,\dots,T\}$. If $t=1$ set $\mu_0^N(dx)=\delta_{\{x_0\}}(dx)$ otherwise input an empirical measure $\mu_{t-1}^N(dx)=\tfrac{1}{N}\sum_{i=1}^N\delta_{\{X_{t-1}^i\}}(dx)$. Set $k=1$.}
\item{For $i\in\{1,\dots,N\}$ generate:
\begin{align*}
X_{t-1+k\Delta_l}^i & =  X_{t-1+(k-1)\Delta_l}^i + a\left(X_{t-1+(k-1)\Delta_l}^i,\overline{\xi}_1(X_{t-1+(k-1)\Delta_l}^i,\mu_{t-1+(k-1)\Delta_l}^N)\right) \textcolor{black}{\Delta_l}  + \\ & b\left(X_{t-1+(k-1)\Delta_l}^i,\overline{\xi}_2(X_{t-1+(k-1)\Delta_l}^i,\mu_{t-1+(k-1)\Delta_l}^N)\right) \times \\ &[W_{t-1+k\Delta_l}^i - W_{t-1+(k-1)\Delta_l}^i]
\end{align*}
where
\begin{eqnarray*}
\overline{\xi}_m(X_{t-1+(k-1)\Delta_l}^i,\mu_{t-1+(k-1)\Delta_l}^N) & = & \frac{1}{N}\sum_{j=1}^N \xi_m(X_{t-1+(k-1)\Delta_l}^i,X_{t-1+(k-1)\Delta_l}^j)\quad m\in\{1,2\}\\
\mu_{t-1+(k-1)\Delta_l}^N(dx) & = & \frac{1}{N}\sum_{j=1}^N\delta_{\{X_{t-1+(k-1)\Delta_l}^j\}}(dx) \\
\left[W_{t-1+k\Delta_l}^i - W_{t-1+(k-1)\Delta_l}^i\right] & \stackrel{\textrm{ind}}{\sim} & \mathcal{N}_{d}(0,\Delta_l I_d).
\end{eqnarray*}
Set $k=k+1$, if $k=\Delta_l^{-1}+1$ go to step 3.~otherwise go to the start of step 2..}
\item{Output all the required laws $\mu_{t-1+\Delta_l}^N,\dots,\mu_{t}^N$.}
\end{enumerate}
\caption{Approximating the Laws when starting with a particle approximation at time $t-1$, $t\in\mathbb{N}$.}
\label{alg:basic_method}
\end{algorithm}

Let $l_*\in\mathbb{N}_0$ be given and $\{N_l\}_{\textcolor{black}{l \geq l_*}}$,  be an increasing 
sequence of non-negative integers such that $\lim_{l\rightarrow\infty} N_l=\infty$.  
Our objective will be first to generate Algorithm \ref{alg:basic_method} sequentially in $t$ for $l=l_*$ and $N_{l_*}$ samples.  Then using the particle system that we have generated,
to \emph{plug-in all the required laws} to a simulation of the Markov kernel $P^l_{\mu_{t-1}^{l,N_l},t}$;  we denote the associated
stationary distribution of $P^l_{\mu_{t-1}^{l,N_l},t}$  as $\Pi^l$.  Note that in effect here,  one runs
Algorithm \ref{alg:basic_method} independently of $P^l_{\mu_{t-1}^{l,N_l},t}$ and the latter kernel,  conditional
on the output of Algorithm \ref{alg:basic_method} is a conventional Euler-Maruyama method.
Subsequently for $l>l_{*}$,  we will be running Algorithm \ref{alg:basic_method_coup} sequentially in $t$
with $N_l$ and $N_{l-1}$ samples
and then simulating a coupling of $P^l_{\mu_{t-1}^{l,N_l},t}$ and $P^{l-1}_{\widetilde{\mu}_{t-1}^{l-1,N_{l-1}},t}$
conditional on the simulation of Algorithm \ref{alg:basic_method_coup},  that is,  \emph{plugging-in all the required laws} and generating Algorithm \ref{alg:basic_method_coup} independently of all other randomness.  The 
stationary distribution
of $P^l_{\mu_{t-1}^{l,N_l},t}$ is still $\Pi^l$.

\begin{algorithm}[h!]
\begin{enumerate}
\item{Input $l\in\mathbb{N}_0$ the level of discretization, $(N_l,N_{l-1})\in\mathbb{N}^2$, $N_{l-1}<N_l$,  the number of particles,  $t\in\{1,\dots,T\}$. If $t=1$ set $\mu_0^{l,N_l}(dx)=\widetilde{\mu}_0^{l-1,N_{l-1}}(dx)=\delta_{\{x_0\}}(dx)$ otherwise input a pair of empirical measures $\mu_{t-1}^{l,N_l}(dx)=\tfrac{1}{N_l}\sum_{i=1}^{N_l}\delta_{\{X_{t-1}^{l,i}\}}(dx)$, 
$\widetilde{\mu}_{t-1}^{l-1,N_{l-1}}(dx)=\tfrac{1}{N_{l-1}}\sum_{i=1}^{N_{l-1}}\delta_{\{\widetilde{X}_{t-1}^{l-1,i}\}}(dx)$.  Set $k=1$.}
\item{For $i\in\{1,\dots,N_l\}$ generate:
\begin{align*}
X_{t-1+k\Delta_l}^{l,i} & =  X_{t-1+(k-1)\Delta_l}^{l,i} + a\left(X_{t-1+(k-1)\Delta_l}^{l,i},\overline{\xi}_1(X_{t-1+(k-1)\Delta_l}^{l,i},\mu_{t-1+(k-1)\Delta_l}^{l,N_l})\right) \textcolor{black}{\Delta_l}  + \\ & b\left(X_{t-1+(k-1)\Delta_l}^{l,i},\overline{\xi}_2(X_{t-1+(k-1)\Delta_l}^{l,i},\mu_{t-1+(k-1)\Delta_l}^{l,N_l})\right)\times \\ &[W_{t-1+k\Delta_l}^i - W_{t-1+(k-1)\Delta_l}^i]
\end{align*}
where
\begin{eqnarray*}
\overline{\xi}_m(X_{t-1+(k-1)\Delta_l}^i,\mu_{t-1+(k-1)\Delta_l}^{l,N_l}) & = & \frac{1}{N_l}\sum_{j=1}^{N_l} \xi_m(X_{t-1+(k-1)\Delta_l}^{l,i},X_{t-1+(k-1)\Delta_l}^{l,j})\quad m\in\{1,2\}\\
\mu_{t-1+(k-1)\Delta_l}^{l,N_l}(dx) & = & \frac{1}{N_l}\sum_{j=1}^{N_l}\delta_{\{X_{t-1+(k-1)\Delta_l}^{l,j}\}}(dx).
\end{eqnarray*}
Set $k=k+1$, if $k=\Delta_l^{-1}+1$ go to step 3.~otherwise go to the start of step 2..}
\item{For $i\in\{1,\dots,N_{l-1}\}$ compute $\widetilde{X}_{t-1+k\Delta_{l-1}}^{l-1,i}$, and $\overline{\xi}_m(\tilde{X}_{t-1+(k-1)\Delta_l}^i,\tilde{\mu}_{t-1+(k-1)\Delta_l}^{l,N_l})$ similarly as above.
Set $k=k+1$, if $k=\Delta_{l-1}^{-1}+1$ go to step 4.~otherwise go to the start of step 3..}
\item{Output all the required laws $\mu_{t-1+\Delta_l}^{l,N_l},\dots,\mu_{t}^{l,N_l}$,  $\widetilde{\mu}_{t-1+\Delta_l}^{l-1,N_{l-1}},\dots,\widetilde{\mu}_{t}^{l-1,N_{l-1}}$.}
\end{enumerate}
\caption{Approximating the Consecutive Laws when starting  at time $t-1$, $t\in\mathbb{N}$}
\label{alg:basic_method_coup}
\end{algorithm}

\subsection{Overall Strategy}

We now present our main strategy as given in \cite{ub_pf}; see also \cite{chada,langevin} for a related approaches,  which are not easily adapted to this context.
We suppose that $\varphi:\mathbb{R}^d\rightarrow\mathbb{R}$ is a functional of interest and that for every 
$l\in\mathbb{N}_0$,  $\Pi^l(\varphi) = \int_{\mathbb{R}^d}\varphi(x)\Pi^l(dx)$ is finite.
 Let $\mathbb{P}_L$ be any positive probability mass function on $\mathbb{N}_{l_*}:=\{l_*,l_*+1,\dots\}$. Let $\{\xi_l\}_{l\in \mathbb{N}_{l_*}}$
be any sequence of independent random variables, such that
\begin{eqnarray*}
\mathbb{E}[\xi_{l_*}] & = & \mathbb{E}^{P}[\Pi^{l_*}(\varphi)] \\
\mathbb{E}[\xi_{l}] & = & \mathbb{E}^{P}[\Pi^{l}(\varphi) - \Pi^{l-1}(\varphi)] =: \mathbb{E}^{P}[[\Pi^l-\Pi^{l-1}](\varphi)] \quad l\in\{l_*+1,l_*+2,\dots\},
\end{eqnarray*}
where $\mathbb{E}^{P}$ is the expectation w.r.t.~the law associated to the simulated systems in 
Algorithm \ref{alg:basic_method} and Algorithm \ref{alg:basic_method_coup}.
Now, let $L$ be a random variable with probability $\mathbb{P}_L$ that is independent of the sequence $\{\xi_l\}_{l\in \mathbb{N}_{l_*}}$ then
\begin{align}
\label{eq:single_term_est}
\widehat{\pi}(\varphi) = \frac{\xi_{\textcolor{black}{L}}}{\mathbb{P}_L(\textcolor{black}{L})},
\end{align}
will be shown to be an unbiased estimator of $\pi(\varphi)$; see \cite{mcl,rhee} for the initial statement and proof. Note however, that we have an additional level of complexity than \textcolor{black}{the above-mentioned papers}, as $\Pi^l$ are random measures.
Moreover, if the following conditions hold,
\begin{equation}\label{eq:finite_var}
\sum_{l\in\mathbb{N}_{l_*}}\frac{\mathbb{E}[\xi_l^2]}{\mathbb{P}_L(l)} < +\infty,
\end{equation}
then the estimator $\widehat{\pi}(\varphi)$ has finite variance. There is also the independent sum-estimator, which can be better than this estimator and has been described in \cite{rhee}.  The main challenge is then to construct the sequence $\{\xi_l\}_{l\in \mathbb{N}_{l_*}}$. 

Typically, one will run $M \in \mathbb{N}$ independent replicates of \eqref{eq:single_term_est} and use the average
\begin{equation}\label{eq:pi_avrg}
\widehat{\pi}(\varphi)_{\text{avg}} := \frac{1}{M}\sum_{i=1}^M \widehat{\pi}(\varphi)^{i},
\end{equation}
where $\widehat{\pi}(\varphi)^{i}$ represents the $i$-th independent replicate of the estimate.

To continue with our discussion we will need a positive probability mass-function $\mathbb{P}_p$
on $\mathbb{N}_0$ and a sequence of non-decreasing,  non-negative integers $\{I_p\}_{p\in\mathbb{N}_0}$
with $\lim_{p\rightarrow\infty}I_p=\infty$.

\subsubsection{Approximation of $\Pi^{l_*}(\varphi)$}\label{sec:ub_mcmc}

Throughout the section $l\in\mathbb{N}_{l_*}$ is fixed.  Our method for constructing $\xi_{l_*}$ is detailed in Algorithm \ref{alg:base_level}.

\begin{algorithm}[h!]
\begin{enumerate}
\item{Input $N_{l_*}$.}
\item{Generate $P\sim\mathbb{P}_{\textcolor{black}{P}}$}
\item{Generate Algorithm \ref{alg:basic_method}  with $N_{l_*}$ particles,  sequentially until time $I_{\textcolor{black}{P}}$ where the empirical measures
at any time $t\in\{1,\dots,I_{\textcolor{black}{P}}\}$ have been obtained from time $t-1$ and the case $t=0$ has been specified
in Algorithm \ref{alg:basic_method} .}
\item{For $t\in\{1,\dots,I_P\}$ generate $U_{t}^{l_{*}}|u_{t-1}^{l_{*}}$ using 
$P_{\mu,t-1}^{l_*}(u_{t-1}^{l_{*}},\cdot)$ where $\mu=\mu_{t-1}^{l_*,N_{l_*}}$, all of the laws 
$\mu_{t-1}^{l_*,N_{l_*}},\mu_{t-1+\Delta_{l_*}}^{l_*,N_{l_*}}, \dots, \mu_{t-\Delta_{l_*}}^{l_*,N_{l_*}}$
needed are obtained in Step 3.~and $u_{0}^{l_{*}}=x_0$.}
\item{If ${\textcolor{black}{P}}=0$ return
$$
\xi_{l_*} = \frac{1}{\mathbb{P}_{\textcolor{black}{P}}({\textcolor{black}{p}})}\frac{1}{I_{\textcolor{black}{P}}}\sum_{t=1}^{I_{\textcolor{black}{P}}}\varphi(u_t^{l_*})
$$
otherwise return 
$$
\xi_{l_*} = \frac{1}{\mathbb{P}_{\textcolor{black}{P}}(\textcolor{black}{p})}
\left\{
\frac{1}{I_{\textcolor{black}{P}}}\sum_{t=1}^{I_{\textcolor{black}{P}}}\varphi(u_t^{l_*}) - 
\frac{1}{I_{{\textcolor{black}{P}}-1}}\sum_{t=1}^{I_{{\textcolor{black}{P}}-1}}\varphi(u_t^{l_*})
\right\}.
$$
}
\end{enumerate}
\caption{Simulation of $\xi_{l_*}$}
\label{alg:base_level}
\end{algorithm}

The approach as developed in Algorithm \ref{alg:base_level} is a simple adaptation of the method in \cite{ub_pf}
in the context here,  except for that method one does not have to feed the empirical measures into any simulation as we have done here.  Note that in practice one would run Step 3.~and Step 4.~concurrently,  that is at each time they are simulated at the same $\Delta_{l_*}$ order increments,  for computational efficiency.  However,  for clarity of presentation we have de-coupled the two steps.  The key property of the estimator,  that will help to ensure that our final estimator is unbiased is that we will show almost surely
\begin{equation}\label{eq:exp_proof}
\mathbb{E}[\xi_{l_*}|\mathscr{L}] = \Pi^{l_{*}}(\varphi).
\end{equation}
where $\mathscr{L}$ is the $\sigma$-algebra generated by Algorithm \ref{alg:basic_method} and Algorithm \ref{alg:basic_method_coup} along with $L$ generated from $\mathbb{P}_L$ (independently of all other random variables. The property in \eqref{eq:exp_proof} is intrinsically based on the convergence of $\mathbb{E}[\frac{1}{I_p}\sum_{t=1}^{I_p}\varphi(u_t^{l_*})|\mathscr{L}]$. The details are in the proof of our main result,  but we try to give some intuition here.

\subsubsection{Approximation of $[\Pi^l-\Pi^{l-1}](\varphi)$}\label{sec:ub_inc_gen}

Our objective is now to provide, for $l\in\{l_*+1,\textcolor{black}{l_*}+2,\dots\}$ fixed, an estimator of $[\Pi^l-\Pi^{l-1}](\varphi)$, 
such that
\begin{equation}\label{eq:cond_ub}
\mathbb{E}[{[\Pi^l-\Pi^{l-1}]}(\varphi)|\mathscr{L}] = [\Pi^l-\Pi^{l-1}](\varphi).
\end{equation}
One could simply use the method outlined above, independently, for $\Pi_l$ and $\Pi_{l-1}$ and independently for each $l\in\{l_*+1,\textcolor{black}{l_*}+2,\dots\}$. However, this is unlikely to provide an estimator that can achieve \eqref{eq:finite_var} and hence the variance of such an approach is infinite and not useful in practice. We therefore present an alternative method.

To describe the simulation of $\xi_l$,  for $l\in\{l_*+1,\textcolor{black}{l_*+2},\dots\}$,  we will need the method given
in Algorithm \ref{alg:coup_kernel}. The algorithm as stated is essentially a synchronous coupling of the simulation
of an Euler-Maruyama time-discretization.  The main difference is that one has to approximate the unknown laws,  which for the purposes of Algorithm \ref{alg:coup_kernel} this is assumed to be given.  Our method for simulating $\xi_l$ is given in Algorithm \ref{alg:gen_level}.  The method in Algorithm \ref{alg:gen_level} helps one to achieve the property \eqref{eq:cond_ub} as will be proved later on.  The essential point is that the differences $
\mathbb{E}[\frac{1}{I_p}\sum_{t=1}^{I_p}\varphi(u_t^{l})|\mathscr{L}]-\Pi^l(\varphi)$ will converge almost surely to zero and this permits the unbiasedness that we need.  The coupling that is achieved in Algorithm \ref{alg:basic_method_coup} and Algorithm \ref{alg:coup_kernel} will help to yield a finite variance estimator.

\begin{algorithm}[h!]
\begin{enumerate}
\item{Input $l\in\{l_*+1,\dots\}$, $t\in\mathbb{N}$,  the empirical laws
$\mu_{t-1}^{l,N_l},\mu_{t-1+\Delta_l}^{l,N_l},\dots,\mu_{t-\Delta_l}^{l,N_l}$,  
$\mu_{t-1}^{l-1,N_{l-1}},\mu_{t-1+\Delta_{l-1}}^{l-1,N_{l-1}},\dots,\mu_{t-\Delta_{l-1}}^{l-1,N_{l-1}}$ and 
$(u_{t-1},\overline{u}_{t-1})\in\mathbb{R}^{2d}$.}
\item{For $k\in\{1,\dots,\Delta_l^{-1}\}$ run the dynamics
\begin{align*}
X_{t-1+k\Delta_l} & = X_{t-1+k\Delta_l} + a\left(X_{t-1+(k-1)\Delta_l},\overline{\xi}_1(X_{t-1+(k-1)\Delta_l},\mu_{t-1+(k-1)\Delta_l}^{l,N_l})\right) + \textcolor{black}{\Delta_l}  \\ & b\left(X_{t-1+(k-1)\Delta_l},\overline{\xi}_2(X_{t-1+(k-1)\Delta_l},\mu_{t-1+(k-1)\Delta_l}^{l,N_l})\right)\times \\&[W_{t-1+k\Delta_l} - W_{t-1+(k-1)\Delta_l}]
\end{align*}
where $X_{t-1}=u_{t-1}$ and for $k\in\{1,\dots,\Delta_l^{-1}\}$, $\left[W_{t-1+k\Delta_l} - W_{t-1+(k-1)\Delta_l}\right]\stackrel{\textrm{ind}}{\sim}\mathcal{N}(0,\Delta_l I_d)$. Set $U_t=x_{t}$.}
\item{For $k\in\{1,\dots,\Delta_{l-1}^{-1}\}$ run the dynamics for $X_{t-1+k\Delta_{l-1}}$ similarly as before. Set $\overline{U}_t=x_{t}$.}
\item{Return $(u_t,\overline{u_t})$.}
\end{enumerate}
\caption{Simulation of a Coupling of $P_{\mu,t}^l(u_{t-1},\cdot)$ and $P_{\overline{\mu},t}^{l-1}(\overline{u}_{t-1},\cdot)$.}
\label{alg:coup_kernel}
\end{algorithm}

\begin{algorithm}[h!]
\begin{enumerate}
\item{Input $l\in\{l_*+1,\dots\}$,  $(N_{l},N_{l-1})$.}
\item{Generate $P\sim\mathbb{P}_{\textcolor{black}{P}}$.}
\item{Generate Algorithm \ref{alg:basic_method_coup}  with $(N_{l},N_{l-1})$ particles,  sequentially until time $I_P$ where the empirical measures
at any time $t\in\{1,\dots,I_P\}$ have been obtained from time $t-1$ and the case $t=0$ has been specified
in Algorithm \ref{alg:basic_method_coup} .}
\item{For $t\in\{1,\dots,I_P\}$ generate $(U_{t}^{l},\overline{U}_t^{l-1})|(u_{t-1}^{l},\overline{u}_{t-1}^{l-1})$ 
from the coupling of $P_{\mu,t}^l(u_{t-1}^l,\cdot)$ and $P_{\overline{\mu},t}^{l-1}(\overline{u}_{t-1}^{l-1},\cdot)$
given in Algorithm \ref{alg:coup_kernel}
where $\mu=\mu_{t-1}^{l,N_{l}}$,  $\overline{\mu}=\mu_{t-1}^{l-1,N_{l-1}}$,   all of the laws 
$\mu_{t-1}^{l,N_{l}},\mu_{t-1+\Delta_{l}}^{l,N_{l}}, \dots, \mu_{t-\Delta_{l}}^{l,N_{l}}$, 
$\mu_{t-1}^{l-1,N_{l-1}},\mu_{t-1+\Delta_{l-1}}^{l,N_{l-1}}, \dots, \mu_{t-\Delta_{l-1}}^{l,N_{l-1}}$,
needed are obtained in Step 3.~and $u_{0}^{l}=\overline{u}_0^{l-1}=x_0$.}
\item{If $\textcolor{black}{P}=0$ return
$$
\xi_{l} = \frac{1}{\mathbb{P}_{\textcolor{black}{P}}(\textcolor{black}{p})}
\left\{
\frac{1}{I_{\textcolor{black}{P}}}\sum_{t=1}^{I_{\textcolor{black}{P}}}\varphi(u_t^{l})
-
\frac{1}{I_{\textcolor{black}{P}}}\sum_{t=1}^{I_{\textcolor{black}{P}}}\varphi(\overline{u}_t^{l-1})
\right\}
$$
otherwise return $\xi_{l}$ as 
$$
\frac{1}{\mathbb{P}_{\textcolor{black}{P}}(\textcolor{black}{p})}
\left[\frac{1}{I_{\textcolor{black}{P}}}\sum_{t=1}^{I_{\textcolor{black}{P}}}\varphi(u_t^{l}) - 
\frac{1}{I_{\textcolor{black}{P}}}\sum_{t=1}^{I_{\textcolor{black}{P}}}\varphi(\overline{u}_t^{l-1})
\right] - 
\left[\frac{1}{I_{\textcolor{black}{P}-1}}\sum_{t=1}^{I_{\textcolor{black}{P}-1}}\varphi(u_t^{l}) - 
\frac{1}{I_{{\textcolor{black}{P}}-1}}\sum_{t=1}^{I_{\textcolor{black}{P}-1}}\varphi(\overline{u}_t^{l-1})
\right].
$$
}
\end{enumerate}
\caption{Simulation of $\xi_{l}$.}
\label{alg:gen_level}
\end{algorithm}

\subsubsection{Final Methodology and Estimator}

We now consolidate the above discussion by summarizing our proposed methodology to unbiasedly estimate $\pi(\varphi)$ and this is presented in Algorithm \ref{alg:final}. As implied by \eqref{eq:pi_avrg} Algorithm \ref{alg:final} can be run $M-$times on parallel.  The choice of $\{N_l\}_{l\geq l_*}$,  $\{I_P\}_{P\in\mathbb{N}_0}$,  $\mathbb{P}_L$
and $\mathbb{P}_P$ is discussed in Section \ref{sec:theory}.

The approach that we have considered as stated previously,  follows that of \cite{ub_pf} but as also mentioned,
there are alternatives based on \cite{chada, langevin}.   These  previous papers are rather dependent upon the notion that the simulated Markov kernel (i.e.~the $P^l_{\mu,t}$ in our notation) is time-homogenous.  This is critical when adopting the methodology of \cite{glynn2} which those works use and as is clear from our context we do not have this property.  Therefore we have concentrated upon the ideas in \cite{ub_pf}.

An alternative idea is to use a double randomization that focusses upon the systems generated in Algorithms
\ref{alg:basic_method} and \ref{alg:basic_method_coup}.  In principle we expect that it is possible to do this,  but we expect that the resulting mathematical analysis is more complicated and the addition of an extra chain (i.e.~the approaches in
Algorithms \ref{alg:base_level} and \ref{alg:gen_level}) does not add a significant cost versus using Algorithms
\ref{alg:basic_method} and \ref{alg:basic_method_coup} on their own; hence we have proceeded with Algorithm \ref{alg:final}.

\begin{algorithm}[h!]
\caption{Unbiased estimator $\widehat{\pi(\varphi)}$.}
\label{alg:final}
{\bf Input}: $\mathbb{P}_L$.
\begin{enumerate}
\item{Sample $L\sim\mathbb{P}_L$.}
\item{If $L=l_*$,  generate $\xi_{l_*}$ using Algorithm \ref{alg:base_level}
and return 
$$
\widehat{\pi}(\varphi) = \frac{\xi_{l_*}}{\mathbb{P}_L(l_*)}.
$$
}
\item{If $L>l_*$,  generate
$\xi_{l}$ using Algorithm \ref{alg:gen_level}
and return 
$$
\widehat{\pi}(\varphi) = \frac{\xi_{l}}{\mathbb{P}_L(l)}.
$$
}
\end{enumerate}
\end{algorithm}

\subsection{\textcolor{black}{Summary of Algorithms}}
\textcolor{black}{For convenience of the reader, we briefly provide an overview of the algorithms we have presented.
The methodology of our unbiased estimator $\widehat{\pi}(\varphi)$, given by \eqref{eq:single_term_est} is provided in Algorithm \ref{alg:final}. We note the theory, related to our unbiased method is provided through Theorem  \ref{thm:2nd}  and Corollary \ref{cor:main_cor}. 
To compute the unbiased estimator we require the simulation of either $\xi_{l_*}$ (which approximates $\Pi^{l_*}(\varphi)$) or  $\xi_{l}$ (which approximates $[\Pi^{l}-\Pi^{l-1}](\varphi)$) . These are provided in Algorithm \ref{alg:base_level} or \ref{alg:gen_level}. 
For the quantity $\xi_l$ we need the computation of the couplings $P^l_{\mu,t}$ which is discussed in in Algorithm \ref{alg:coup_kernel}. 
Finally Algorithm \ref{alg:basic_method} provides a Monte Carlo approximation of the law $\mu_t^l$, while this is extended to Algorithm \ref{alg:basic_method_coup}
 when we require to sample from a dependent coupling of a pair of laws $(\mu^l_t,\mu^{l-1}_t)$. In other words, the first two algorithms consider the time-dscretization of the MVSDEs, based on the discretization \eqref{eq:euler}.
}

\newpage

\section{Theoretical Results}
\label{sec:theory}

\subsection{Notation}

Denote by $\mathcal{C}_b(\mathbb{R}^{d_1},\mathbb{R}^{d_2})$ the set of $\mathbb{R}^{d_2}$ valued bounded continuous functions whose domain is $\mathbb{R}^{d_1}$ and equip it with the norm $\|f\| = \sup_{x\in\mathbb{R}^{d_1}}|f(x)|$. Denote by $\mathcal{C}^1_b(\mathbb{R}^{d_1},\mathbb{R}^{d_2})$ the set of continuously differentiable functions with domain $\mathbb{R}^{d_1}$ and values in $\mathbb{R}^{d_2}$ whose partial derivatives of order $1$ are bounded functions. For a function
For $f\in\mathcal{C}^1_b(\mathbb{R}^{d_1},\mathbb{R}^{d_2})$ define the seminorm $|f|_1 = \max_{i\in\{1,\dots,d_1\}}\sup_{x\in\mathbb{R}^{d_1}} |\partial_{x_i}f(x)|$ and for $f\in\mathcal{C}_b(\mathbb{R}^{d_1},\mathbb{R}^{d_2})\cap \mathcal{C}^1_b(\mathbb{R}^{d_1},\mathbb{R}^{d_2})$ define the norm $\|f\|_1=\max(\|f\|,|f|_1)$. Denote by $\mathcal{C}^{\textrm{Lip}}(\mathbb{R}^{d_1},\mathbb{R}^{d_2})$ the set of Lipschitz continuous functions $f:\mathbb{R}^{d_1}\to\mathbb{R}^{d_2}$. Define the seminorm $|f|_{\textrm{Lip}}$ and the norm $\|f\|_{\textrm{Lip}}$ for $f\in\mathcal{C}^{\textrm{Lip}}_b(\mathbb{R}^{d_1},\mathbb{R}^{d_2})$ by
$$
|f|_{\mathrm{Lip}} := \sup_{x\not=y} \frac{|f(x)-f(y)|}{|x-y|}, \quad \|f\|_{\textrm{Lip}}:=\max(\|f\|,|f|_{\mathrm{Lip}}).
$$
For the function $a$ denote by $\nabla_1 a(x,y)$ the gradient of the function $x\mapsto a(x,y)$ and by $\nabla_2 a(x,y)$ the gradient of the function $y\mapsto a(x,y)$, similarly for the functions $b,\xi_1,\xi_2$. 

\subsection{Assumptions}

In order for us to proceed we require a number of assumptions for our theory. We state the following assumptions.
\begin{hypA}\label{assump:A1}
    The functions $a\in\mathcal{C}^1_b(\mathbb{R}^d\times\mathbb{R},\mathbb{R}^d)$, $b\in\mathcal{C}^{\mathrm{Lip}}(\mathbb{R}^d\times\mathbb{R},\mathbb{R}^{d\times d})$, $\xi_1,\xi_2\in\mathcal{C}^{\mathrm{Lip}}(\mathbb{R}^d\times\mathbb{R}^d,\mathbb{R})\cap \mathcal{C}_b(\mathbb{R}^d\times\mathbb{R}^d,\mathbb{R})$.
\end{hypA}
\begin{hypA}\label{assump:A2}
 The following inequality holds
 $$-\sup_{x\in\mathbb{R}^{d+1}}\sup_{|y|=1} y^{\top}\nabla_1 a(x)y >  2\|\nabla_2 a\|\|\xi_1\|_{\mathrm{Lip}} + 2|b|_{\mathrm{Lip}}^2(1+\|\xi_2\|_{\mathrm{Lip}})^2.$$
\end{hypA}

\textcolor{black}{To state our final assumption we use some additional notation to shorten the subsequent discussion.   If $l=l_*$
$$
 \overline{\xi}_{l,P}  =
\begin{cases}
\frac{1}{I_{P}}\sum_{t=1}^{I_{P}}\varphi(u_t^{l})
& \textrm{if}~P=0\\
\left\{
\frac{1}{I_{P}}\sum_{t=1}^{I_{P}}\varphi(u_t^{l}) - 
\frac{1}{I_{P-1}}\sum_{t=1}^{I_{P-1}}\varphi(u_t^{l})
\right\}
& \textrm{if}~P>0
 \end{cases} 
$$
and for $l\in\{l_*+,\dots\}$, we have that 
$$
 \overline{\xi}_{l,P}  =
\begin{cases}
\frac{1}{I_{P}}\sum_{t=1}^{I_{P}}\varphi(u_t^{l})
- \frac{1}{I_{P}}\sum_{t=1}^{I_{P}}\varphi(\overline{u}_t^{l-1}) \\
\left[\frac{1}{I_{P}}\sum_{t=1}^{I_{P}}\varphi(u_t^{l}) - 
\frac{1}{I_{P}}\sum_{t=1}^{I_{P}}\varphi(\overline{u}_t^{l-1})
\right] - 
\left[\frac{1}{I_{P-1}}\sum_{t=1}^{I_{P-1}}\varphi(u_t^{l}) - 
\frac{1}{I_{P-1}}\sum_{t=1}^{I_{P-1}}\varphi(\overline{u}_t^{l-1})
\right],
 \end{cases} 
$$
where $ \overline{\xi}_{l,P}$ is $\xi_l\mathbb{P}_{\mathsf{P}}(P)$ as given in Algorithms \ref{alg:base_level} and \ref{alg:gen_level}, however the addition of the $P$ subscript will make the statement below much easier to follow.
}
\begin{hypA}\label{assump:A3}
\textcolor{black}{For each $l\in\{l_{*},l_{*}+1,\dots\}$ we have
$$
\lim_{P\rightarrow\infty}\frac{\mathbb{E}[|\overline{\xi}_{l,P+1}|]}{\mathbb{E}[|\overline{\xi}_{l,P}|]} <1.
$$
In addition
$$
\lim_{l\rightarrow\infty}\frac{\sum_{P=0}^{\infty}\mathbb{E}[|\overline{\xi}_{l+1,P}|]}
{\sum_{P=0}^{\infty}\mathbb{E}[|\overline{\xi}_{l,P}|]} <1.
$$}
\end{hypA}

Let us now discuss the importance of each assumption above.
It is well known that under condition \hyperref[assump:A1]{(A1)} a unique strong solution exists for the SDE \eqref{eq:sde}, with details in \cite{mish}. The assumptions \hyperref[assump:A1]{(A1-2)} are needed to guarantee the existence of the invariant measure for both the continuous McKean-Vlasov SDE and the discretized SDE and to ensure the stability of the Euler Scheme. To show the existence of the invariant measure of the McKean-Vlasov SDE \eqref{eq:sde} we utilize \cite[Theorem 3.1]{wang} which guarantees geometric ergodicity.
which in turn requires one to verify conditions (H1), (H2'), and (H3) of \cite{wang}. 
\textcolor{black}{For \hyperref[assump:A3]{(A3)},  which ensures that our estimator is finite in expectation,  we expect that 
$$
\mathbb{E}[|\overline{\xi}_{l,P+1}|] = \mathcal{O}\left(\Delta_l^{1/2}\{I_P^{-1/2}+N_l^{-1/2}\}\right)
$$
so that indeed this is a reasonable assumption, at least for suitable choices of $I_P$ and $N_l$.  We have not proved the above result, but discuss this point further in Section \ref{sec:theo_disc}.
}

\subsection{Unbiasedness of the Estimator}

Our main result is the unbiasedness of the estimator.  Note that we require $l_{*}$ to be large enough and this is assumed.  The proof of this result depends itself on numerous egrodicity results that are established in Appendix
\ref{app:erg}.

\begin{theorem}\label{theo:ub}
    Assume \hyperref[assump:A1]{(A1-2)}.   Then for any $\varphi\in\mathcal{C}_b(\mathbb{R}^d,\mathbb{R})\cap\mathcal{C}^{\mathrm{Lip}}(\mathbb{R}^d,\mathbb{R})$ we have 
    $$
\mathbb{E}\left[\widehat{\pi}(\varphi)\right] = \pi(\varphi),
$$
\textcolor{black}{for $l_*$ large enough}.
\end{theorem}

\begin{proof}
The proof is completed in several simple computations as given below:
    \begin{align*}
            \mathbb{E}[\widehat{\pi}(\varphi)]&=\mathbb{E}[\textcolor{black}{\frac{1}{\mathbb{P}_P(P)\mathbb{P}_L(L)}}\sum_{\textcolor{black}{L}=1}^{\infty}\sum_{P=1}^{\infty}\Big[\frac{1}{I_P}\sum_{t=1}^{I_P}\varphi(U_t^{\textcolor{black}{L}})\\&-
            \frac{1}{I_{P}}\sum_{t=1}^{I_{P}}\varphi(\overline{U}_t^{\textcolor{black}{L}-1})\Big]-
           \Big [
            \frac{1}{I_{P-1}}\sum_{t=1}^{I_{P-1}}\varphi(U_t^{\textcolor{black}{L}}) -
            \frac{1}{I_{P-1}}\sum_{t=1}^{I_{P-1}}\varphi(\overline{U}_t^{\textcolor{black}{L}-1})\Big]] \\
            &=\lim_{L\to\infty}\lim_{P\to\infty}\mathbb{E}[\frac{1}{I_P}\sum_{t=1}^{I_P}\varphi(U_t^{\textcolor{black}{L}}]\\
            &=\lim_{L\to\infty}\mathbb{E}[\lim_{P\to\infty}\mathbb{E}[\frac{1}{I_P}\sum_{i=1}^{I_P}\varphi(U_t^{\textcolor{black}{L}})\bigg|\mathscr{L}]]\\
            &=\lim_{L\to\infty}\mathbb{E}[\Pi^{L}(\varphi)]\\
            &= \pi(\varphi),
    \end{align*}
    where we have used Theorem \ref{thm:2nd} to go from line three to line four and again
to go the final line  and the interchanges of limits and integrals are justifiable by the bounded convergence theorem.
\end{proof}

\subsection{Discussion}
\label{sec:theo_disc}
Theorem \ref{theo:ub} gives us unbiasedness,  but little else to help us choose the parameters in the method.
We conjecture that, based on theory in \cite{siddig,ub_pf}, that 
the variance is upper-bounded by a term that is as below
\begin{equation}\label{eq:var_conj}
\mathcal{O}\left(
\sum_{l=l_{*}}^{\infty}\sum_{\textcolor{black}{P}=0}^{\infty}\frac{1}{\mathbb{P}_L(l)\mathbb{P}_P(\textcolor{black}{P})}
\left\{
\frac{\Delta_l}{I_{\textcolor{black}{P}}}\left(1+\frac{1}{N_l}\right)
\right\}
\right).
\end{equation}
Note that to achieve this bound,  we expect that we need some ergodicity properties
of the Markov kernel $P_{\mu,t}^l$ with convergence rates that are $l-$independent.  We would expect again that this might only occur under iteration as we have done in this paper.
The expected cost is
$$
\mathcal{O}\left(\sum_{l=l_{*}}^{\infty}\sum_{p=0}^{\infty}
\mathbb{P}_L(l)\mathbb{P}_{\textcolor{black}{P}}(\textcolor{black}{P})
I_{\textcolor{black}{P}}\Delta_l^{-1}N_l^2\right).
$$
In this case it is difficult to choose $\{N_l\}_{l\geq l_*}$,  $\{I_{\textcolor{black}{P}}\}_{\textcolor{black}{P}\in\mathbb{N}_0}$,  $\mathbb{P}_L$
and $\mathbb{P}_P$ so that both  \eqref{eq:var_conj} and the expected cost is finite.
One can choose $N_l=l$, $I_{\textcolor{black}{P}}=2^{\textcolor{black}{P}}$,  $\mathbb{P}_L(l)\propto 2^{-l}(l+1)\log_2(l+2)^2$,  and $\mathbb{P}_P(\textcolor{black}{P})\propto 2^{-\textcolor{black}{P}}(\textcolor{black}{P}+1)\log_2(\textcolor{black}{P}+2)^2$ and the expression in \eqref{eq:var_conj} is finite.

\begin{rem}
We note our ergodicity results assume a discretized MVSDE based on the
Euler-Maruyama discretization. It is important to highlight these results could be potentially extended
to higher-order discretization methods which have more favorables error rates. Examples of this would
include splitting order schemes, which have proven to be successful. 
\end{rem}

\begin{rem}
\textcolor{black}{
By stating $L$ large enough, earlier in the section, we refer to the the fact that one is able to achieve unbiasedness,
in the infinite limit  $L \rightarrow \infty$ \cite{glynn2,rhee}. In practice this is difficult to achieve, as the cost can be infinite, 
so therefore one must truncate the series. Numerically, we would ideally like to choose an sufficient choice, that removes
the bias considerably. This will depend much on the application, and the parameter choices associated to it.}
\end{rem}

\section{Numerical Experiments}
\label{sec:num}
In this section we consider testing our unbiased estimator to approximate the invariant measures for a selection of different MVSDEs. In particular we will consider three models to test, which 
include the Curie-Weiss model, an OU process and a model taking motivation for mathematical neuroscience. We will demonstrate the effectiveness
of our approximation method through different plots such as comparing the MSE to the number of Monte Carlo samples $M$ and approximating the density of the invariant measure. One of our numerical experiments is a parameter estimation problem. Before discussing each model, we provide a brief overview on out simulation setup.

\subsection{Simulation Setting}
For $\textcolor{black}{M\in\mathbb{N}}$ we denote the by $\widehat{\pi}_M$ the estimator described in \eqref{eq:pi_avrg} run with $M$ independent simulations. Let $\varphi:\mathbb{R}\to\mathbb{R}$.
We estimate the mean squared error (MSE) corresponding to $M$ and $\varphi$ we run $50$ independent runs $\{\widehat{\pi}_M^k\}_{k=1}^{50}$ of the method and calculate the MSE given as
$$\textup{MSE} = \frac{1}{50}\sum_{k=1}^{50}(\widehat{\pi}^k_M(\varphi) - \pi(\varphi))^2$$
We set $l_*=3, l_{\mathrm{max}}=10, p_{\mathrm{max}}=7$ and 
\begin{align*}
\mathbb{P}_L(l)&\propto 2^{-l}(l+1)\log(l+2)\mathbb{1}_{\{l_{\mathrm{min}}\leq l\leq l_{\mathrm{max}}\}}, \\
\mathbb{P}_P(p)&\propto 2^{-p}(p+1)\log(p+2)^2\mathbb{1}_{\{0 \leq p \leq p_{\mathrm{max}}\}},
\end{align*}
 and where $N_l=\mathcal{O}(l)$. We denote by $K_h(x)=\mathcal{N}(0,h^2)$ the density of the normal distribution with standard deviation $h$. 
 The density $p$ of $\pi$ is estimated using kernel density estimation (KDE) by
$$p(x) \approx \int_{\mathbb{R}^d} K_h(x-y)~d~\widehat{\pi}_M(y) = \widehat{\pi}_M(K_h(x-\cdot)),$$
for an appropriate choice of $h\in \mathbb{R}$. 
 \textcolor{black}{Furthermore, we verify Assumption \hyperref[assump:A3]{(A3)} for the Curie-Weiss model by estimating the ratios defined in Assumption \hyperref[assump:A3]{(A3)}:
$$Q_1(l) = \lim_{P\rightarrow\infty}\frac{\mathbb{E}[|\overline{\xi}_{l,P+1}|]}{\mathbb{E}[|\overline{\xi}_{l,P}|]},\quad
Q_2(l) = \frac{\sum_{P=0}^{\infty}\mathbb{E}[|\overline{\xi}_{l+1,P}|]}
{\sum_{P=0}^{\infty}\mathbb{E}[|\overline{\xi}_{l,P}|]}.
$$
For each $l\in\{l_*,\dots,l_{\max}\}$ and $P\in\{0,\dots,p_{\max}\}$, we generate $100$ independent realizations $\bar{\xi}_{l,P}^{k}$, $k\in\{1,\dots,100\}$, of $\bar{\xi}_{l,P}$. We construct the estimators
\begin{align}
    \widehat{Q}_1(l) &= \frac{\frac{1}{100}\sum_{k=1}^{100}|\bar{\xi}_{l,p_{\max}}^k|}{\frac{1}{100}\sum_{k=1}^{100}|\bar{\xi}_{l,p_{\max}-1}^k|},&  l\in\{l_*,\dots,l_{\max}\}\label{eq:Q_1_est}\\
\widehat{Q}_2(l) &= \frac{\frac{1}{100}\sum_{p=0}^{p_{\max}}\sum_{k=1}^{100}|\bar{\xi}_{l+1,p}^k|}{\frac{1}{100}\sum_{p=0}^{p_{\max}}\sum_{k=1}^{100}|\bar{\xi}_{l,p}^k|},& l\in\{l_*,\dots,l_{\max}-1\}.\label{eq:Q_2_est}
\end{align}
Assumption \hyperref[assump:A3]{(A3)} requires that both ratios $Q_1$ and $Q_2$ are below $1$, we verify this condition empirically by confirming $\widehat{Q}_1<1$ and $\widehat{Q}_2<1$.}

\subsection{Curie-Weiss Model}
\begin{figure}[h!]
    \centering
    \includegraphics[width=0.4\linewidth]{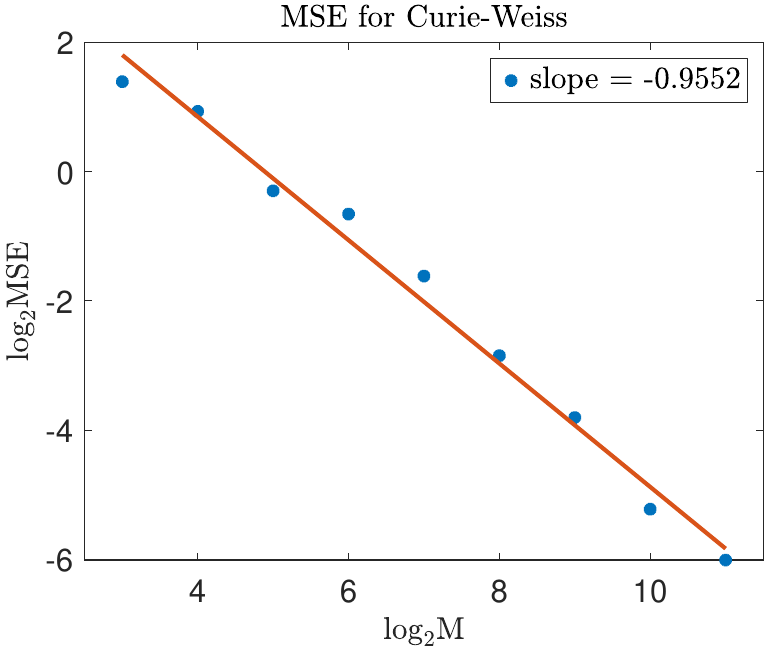}
    \includegraphics[width=0.4\linewidth]{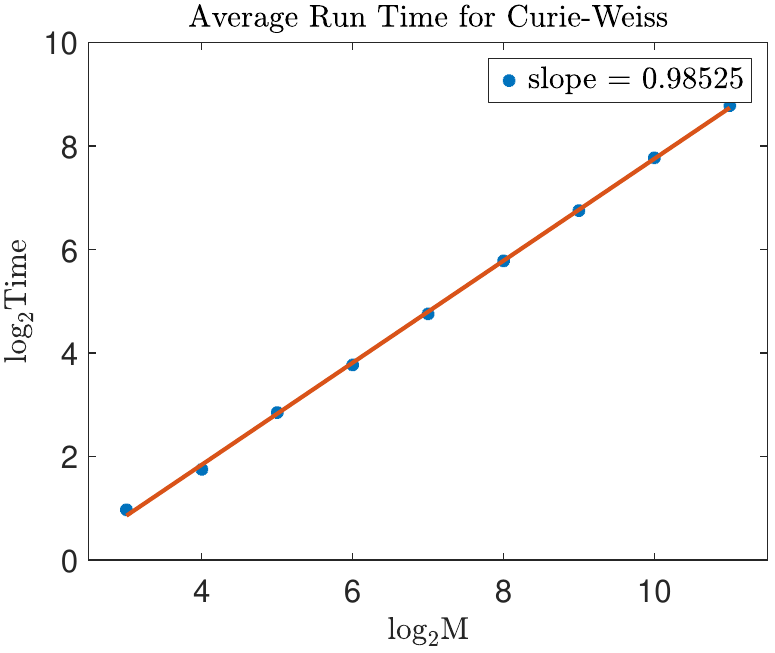}
    \includegraphics[width=0.4\linewidth]{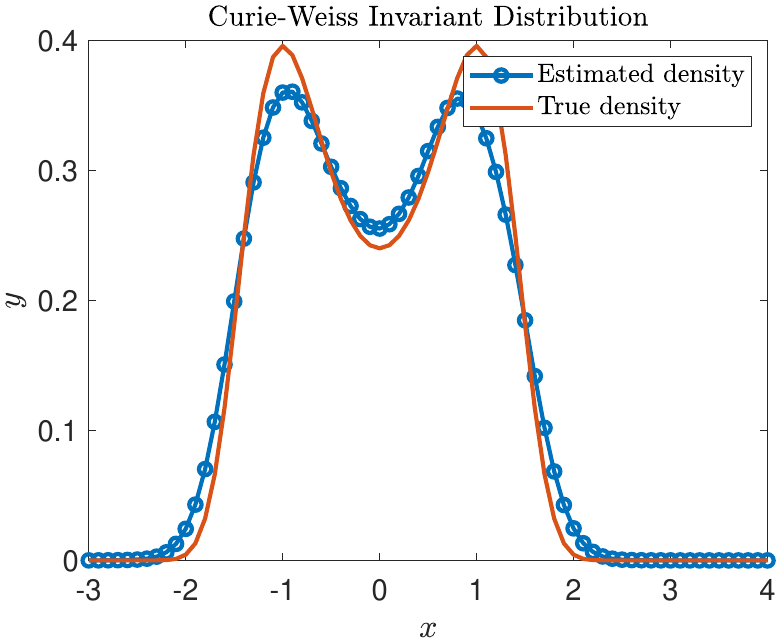}
    \includegraphics[width=0.4\linewidth]{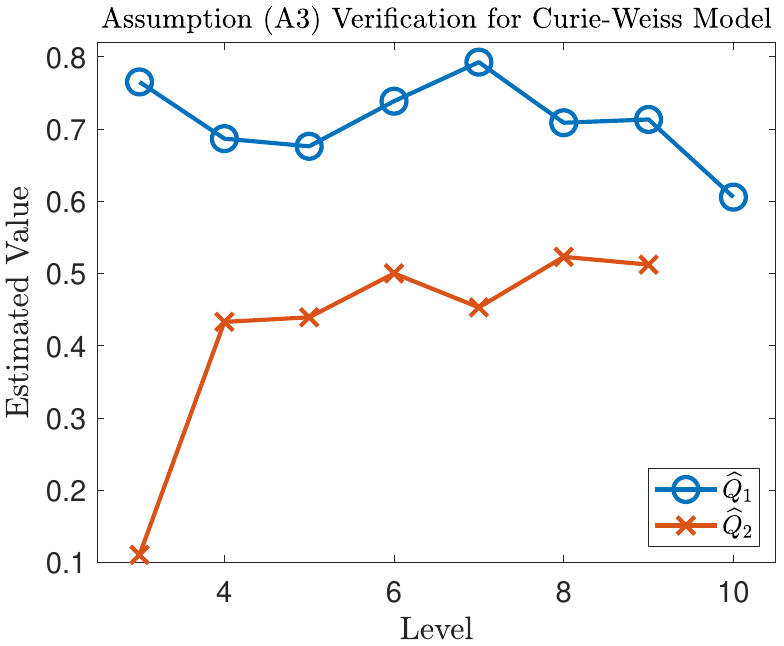}
    \caption{Numerical simulations for the Curie-Weiss model \eqref{eq:cw}. Top left: MSE approximation of $\pi(\varphi)$. Top right: Meeting time for Curie-Weiss model. Bottom left: comparison of exact and approximated invariant distribution. Bottom Right: Numerically verifying Assumption \hyperref[assump:A3]{(A3)} by plotting $\widehat{Q}_1$ and $\widehat{Q}_2$.}
    \label{fig:CW}
\end{figure}

Our first model we test our unbiased methodology on is the following one-dimensional SDE
\begin{equation}
\label{eq:cw}
dX_t = \beta(-X^3_t+X_t + K\mathbb{E}[X_t])dt + \sigma dW_t,
\end{equation}
with the initial condition $X_0=x_0\in\mathbb{R}$, where $\beta,K,\sigma>0$ and $W_t$ is a standard Brownian motion. It is well-known the invariant distribution $\pi$ of this model is absolutely continuous with respect to the Lebesgue measure and has the density
$$p(x) = C\exp\left(-\frac{\beta x^4}{2} + \beta x^2\right),$$
where $C$ is the reciprocal of the normalizing constant. For this example, we set $\beta=1,K=0.25,\sigma=1,x_0=1$ and $\varphi(x)=x^2$. We numerically approximate $C\approx 0.2401$ and $\pi(\varphi)\approx 0.8935$. We approximate $\pi(\varphi)$ using our method to and evaluate $\widehat{\pi}_{M}(\varphi)$. Figure \ref{fig:CW} shows the MSE  $\mathbb{E}[(\widehat{\pi}_{M}(\varphi) - \pi(\varphi))^2]$ and the average running time corresponding to the runs. Furthermore we approximate the density $p$ using our method and KDE. Figure \ref{fig:CW} demonstrates that unbiased estimator works well, as the KDE attains high accuracy of the stationary distribution, and the rates are favourable which are coincide with the discussion on the cost in Section \ref{sec:theory}.  \textcolor{black}{Finally, Figure \ref{fig:CW} presents the estimated ratios $\widehat{Q}_1$ and $\widehat{Q}_2$ defined in \eqref{eq:Q_1_est} and \eqref{eq:Q_2_est}. All values lie strictly below $1$, verifying the bounds required by Assumption \hyperref[assump:A3]{(A3)}.}

\subsection{Parameter Estimation}
Let $\{p_{\theta}(x,y)\}_{\theta\in\Theta}$ be a parameterized family of probability densities on $\mathbb{R}^{d_x}\times \mathbb{R}^{d_y}$. Given a $y\in\mathbb{R}^d$ consider the problem of finding the Maximum Likelihood Estimator (MLE) $$\theta^* = \textup{argmax}_{\theta\in\Theta} p_{\theta}(y) = \textup{argmax}_{\theta\in\Theta} \int p_{\theta}(x,y) dx.$$ 
\cite{kuntz} shows that the MLE is the limit $\theta^* = \lim_{t\to\infty} \theta_t$ with $\theta_t$ being the solution of the following McKean-Vlasov SDE
\begin{equation}\label{eq:kuntz_1}
    \begin{dcases}
        d\theta_t = \left(\int_{\mathbb{R}^d} \nabla_{\theta}\log p_{\theta_t}(x,y) d\mu_t(x)\right)dt,\\
        dX_t = \nabla_x \log p_{\theta_t}(X_t,y)dt + \sqrt{2}dW_t.
    \end{dcases}
\end{equation}
where $\mu_t$ is the law of the process $X_t$ and $W_t$ is a standard Brownian motion. Furthermore, the law $\mu_t$ of the process $X_t$ defined in \eqref{eq:kuntz_1} is absolutely continuous with the Lebesgue measure and $\lim_{t\to\infty} d\mu_t/dx= p_{\theta^*}(\cdot|y)$ where $p_{\theta^*}(x|y)=p_{\theta^*}(x,y)/p_{\theta^*}(y)$ is the posterior. To apply our method we discretize the system as follows: For each $N,l\in\mathbb{N}$ we consider
\begin{equation}\label{eq:param_est_1}
    \begin{dcases}
        \theta_{(k+1)\Delta_l} = \theta_{k\Delta_l} + \left(\int _{\mathbb{R}^d}\nabla_{\theta}\log p_{\theta_{k\Delta_l}}(x,y) d\mu_{k\Delta_l}^N(x)\right)\Delta_l + \Delta_l(B_{(k+1)\Delta_l} - B_{k\Delta_l}),\\
        X^i_{(k+1)\Delta_l} = X^i_{k\Delta_l} + \nabla_x \log p_{\theta_{k\Delta_l}}(X^i_{k\Delta_l},y)\Delta_l + \sqrt{2}(W_{(k+1)\Delta_l} - W_{k\Delta_l}),
    \end{dcases}
\end{equation}
where $\mu_{k\Delta_l}^N = \frac{1}{N}\sum_{i=1}^N\delta_{X_{k\Delta_l}}^i$, $i\in\{1,\dots,N\}$, $k\in\Delta_l^{-1}\mathbb{N}_0$, and $B_t$ is a standard Brownian motion independent of $W_t$.\\

\begin{figure}[h!]
    \centering
    \includegraphics[width=0.4\linewidth]{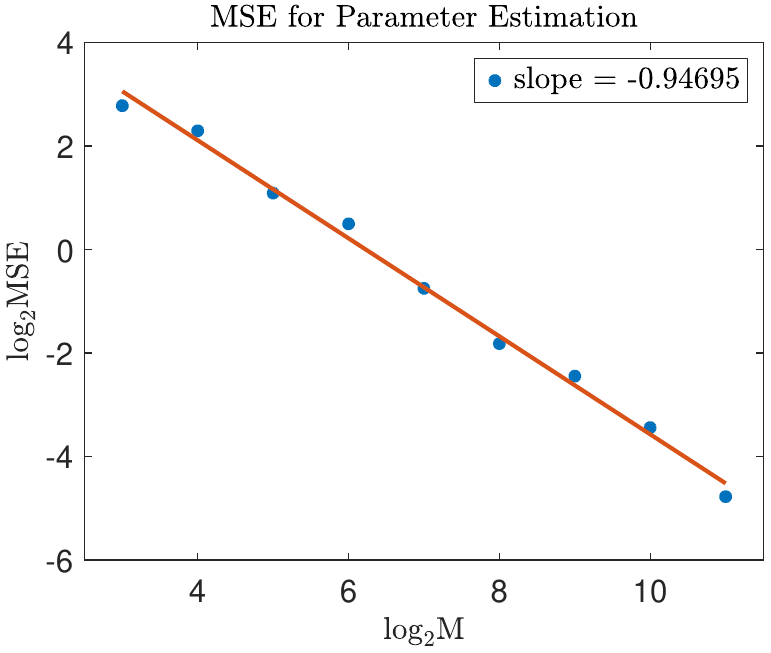}    
    \includegraphics[width=0.4\linewidth]{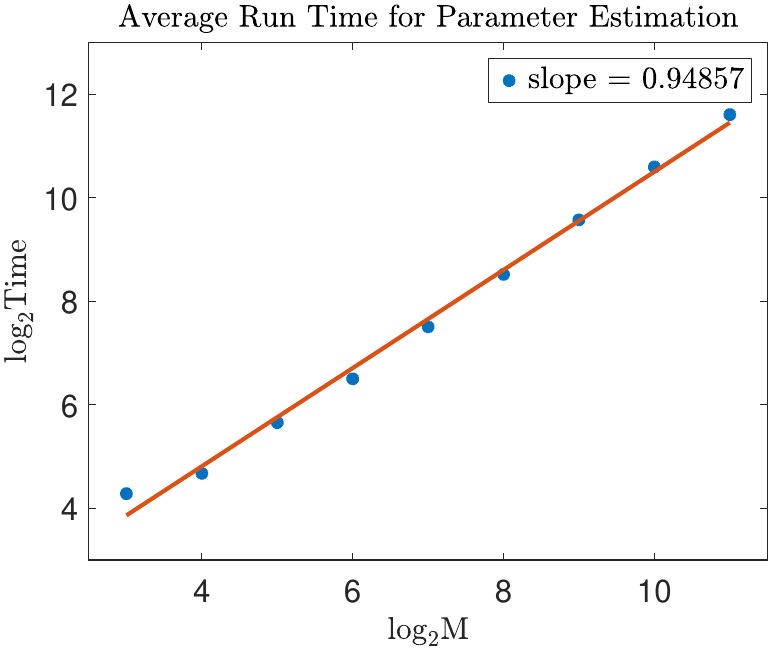}
    \includegraphics[width=0.4\linewidth]{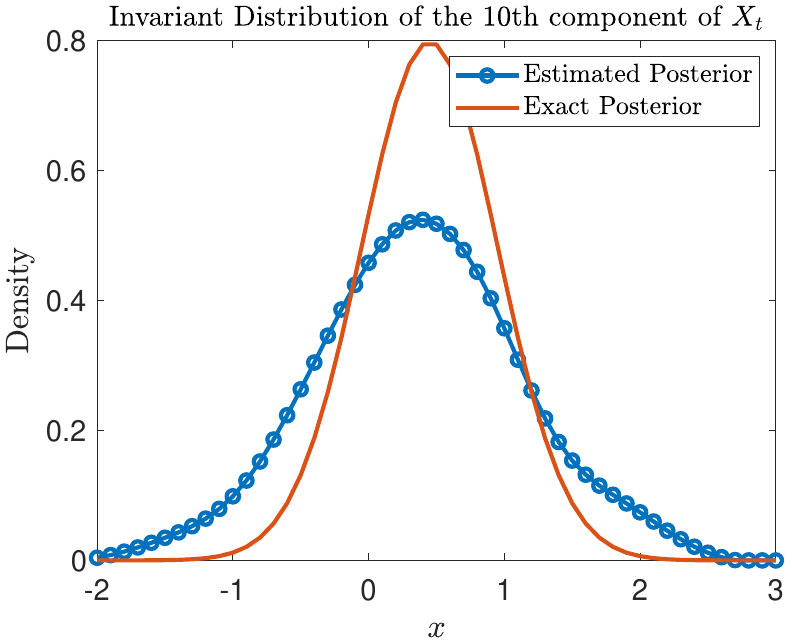}
    \caption{Numerical simulations for parameter estimation example \eqref{eq:kuntz_1}. Top left: MSE approximation of $\phi$. Top right: Meeting time for parameter estimation example. Bottom: comparison of exact and approximated posterior distribution. }
    \label{fig:kuntz}
\end{figure}

\noindent We consider the toy example considered in \cite{kuntz}. For $\theta\in\mathbb{R}$ let $\bar{\theta} \in\mathbb{R}^d$ be the vector whose all components are equal to $\theta$. Let $p_{\theta}(x,y) = \mathcal{N}(y;x,I_{d})\mathcal{N}(x;\bar{\theta},I_{d})$. Let $y=(y_1,\dots,y_d)\in\mathbb{R}^d$, the likelihood $p_{\theta}(y) = \mathcal{N}(y;\bar{\theta},2I_d)$, the posterior $p_{\theta}(x|y)=\mathcal{N}(x;\frac{y+\bar{\theta}}{2},\frac{1}{2}I_d)$, and the MLE has the closed form $\theta^*=\frac{1}{d}\sum_{i=1}^dy_i$. We set $d=10$ and apply our method to the vector $(\theta_t,X_t)$. Denote by $\widehat{\pi}$ the estimate our method returns for the invariant distribution of $(\theta_t,X_t)$ and define the function $\varphi:(\theta,x)\mapsto \theta$. Our estimate of the MLE $\theta^*$ is $\widehat{\pi}(\varphi)$. Figure \ref{fig:kuntz} shows the convergences rate as function of the number of independent samples $M$ and the average running time. Figure \ref{fig:kuntz} shows the estimated posterior of the $10$th component of the process $X_t$ in \eqref{eq:param_est_1}. The results obtained for this experiment, follow similarly to that for the Currie-Weiss Model, where we obtain two rate which are approximately -1 and 1, when comparing $M$ to the MSE and the average run time. By average run time we mean the total sum of all the $M$ runs.

\subsection{3D Neuron Model}
\begin{figure}[h!]
    \centering
    \includegraphics[width=0.4\linewidth]{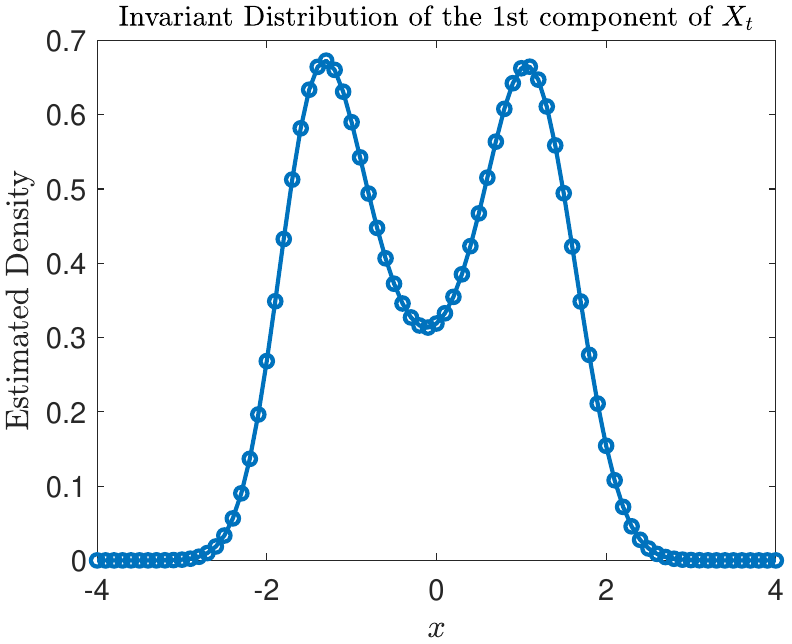}    
    \includegraphics[width=0.4\linewidth]{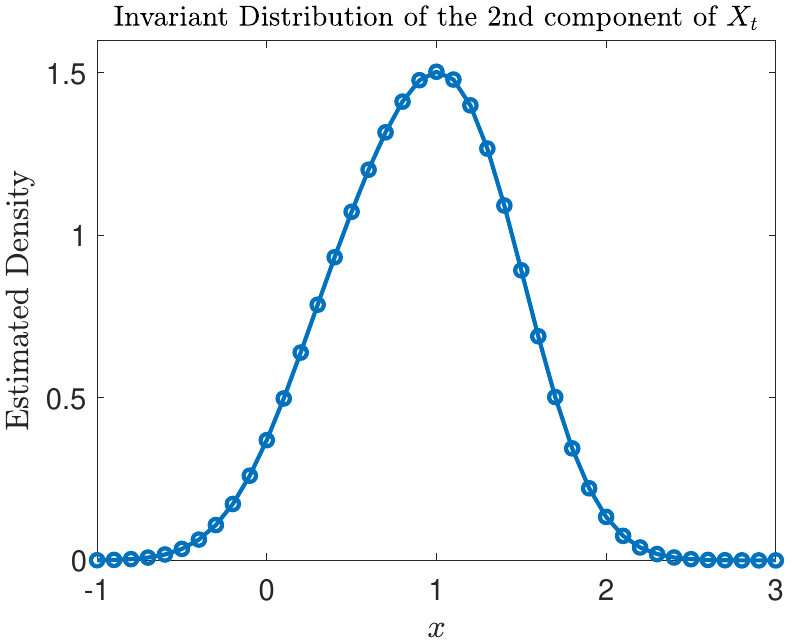}
    \includegraphics[width=0.4\linewidth]{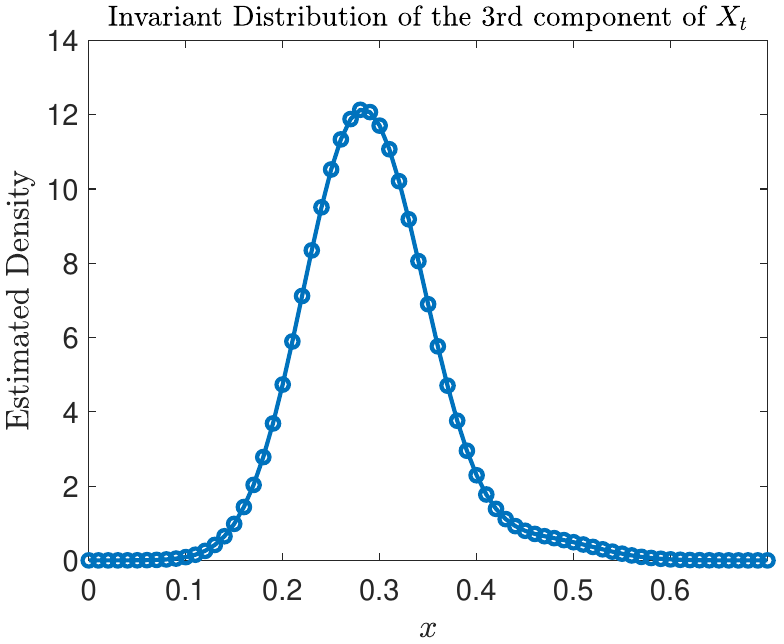}
    \includegraphics[width=0.4\linewidth]{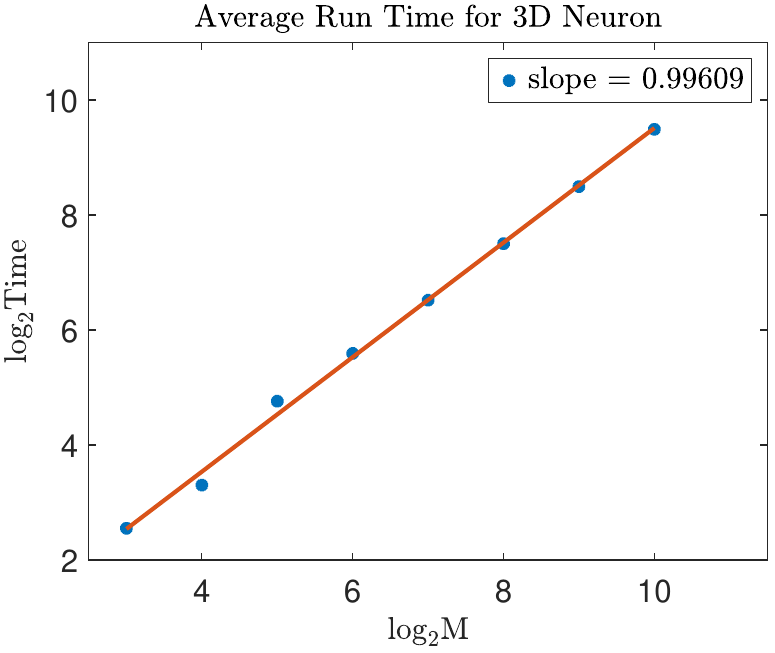}
    \caption{Numerical simulations for the 3D neuron model \eqref{eq:3d}. Top left: approximated Invariant distribution for 1st component. Top right: approximated Invariant distribution for 2nd component. Bottom left: approximated Invariant distribution for 3rd component. Bottom right: Meeting time for parameter estimation example.}
    \label{fig:neuron}
\end{figure}

\textcolor{black}{Our final model we test is inspired from the work of \cite{baladron}, which develops a non globally Lipschitz MV-SDE to model neuron
activity. It is a 3D neuron model which has a specific form of for the drift term and diffusion coefficient. For this model we assume the SDE takes the general form,}
\begin{equation}
\label{eq:3d}
dX_t = a( t, x, \mu) dt +  b( t, x, \mu) dW_t,
\end{equation}
where the drift term, and diffusion coefficient, have the following representation
\begin{align*}
a \left( t, x, \mu \right) 
&:= \left( \begin{array}{c}
x_1 - (x_1)^3 / 3 - x_2 + I - \int_{\R^3} J \left( x_1 - V_{rev} \right) z_3 d \mu (z) \\
c \left( x_1 + a - b x_2 \right) \\
a_r \frac{ T_{max} (1 - x_3) }{ 1 + \exp ( - \lambda ( x_1 - V_T)) } - a_d x_3
\end{array} \right) \\
b \left( t, x, \mu \right) 
&:= \left( \begin{array}{ccc}
b_{ext} & 0 & - \int_{\R^3} b_J \left( x_1 - V_{rev} \right) z_3 d \mu (z) \\
0 & 0 & 0 \\
0 & b_{32} (x) & 0
\end{array} \right)
\end{align*}
with
\begin{equation*}
b_{32} (x) := \mathbb{1}_{\{x_{3} \in (0,1)\}} \sqrt{ a_r \frac{ T_{max} (1 - x_3) }{ 1 + \exp ( - \lambda ( x_1 - V_T)) } + a_d x_3 } \ \Gamma \exp ( - \Lambda / (1 - (2x_3 - 1)^2)),
\end{equation*}
where$T=2$ is chosen as the final time. Finally we set an initial condition and parameter values as
\begin{equation*}
X_0 \sim \mathcal{N} \left( \left( \begin{array}{c} V_0 \\ w_0 \\ y_0 \end{array} \right) , \left( \begin{array}{ccc} \sigma_{V_0} & 0 & 0 \\ 0 & \sigma_{w_0} & 0 \\ 0 & 0 & \sigma_{y_0} \end{array} \right) \right),
\end{equation*}
where the parameters have the values
\begin{equation*}
\begin{array}{ccccccc}
V_0 = 0 & \sigma_{V_0} = 0.4   & a = 0.7   & b = 0.8    & c=0.08   & I = 0.5  & b_{ext} = 0.5 
\\
w_0 = 0.5 & \sigma_{w_0} = 0.4 & V_{rev} = 1   & a_r = 1  & a_d = 1   & T_{max} = 1   & \lambda = 0.2  
\\
y_0 = 0.3 & \sigma_{y_0} = 0.05  & J = 1   & b_J = 0.2  & V_T = 2   & \Gamma = 0.1   & \Lambda = 0.5.
\end{array}
\end{equation*}
Figure \ref{fig:neuron} shows the estimated marginals of the invariant distribution of the process $X_t$, while also presenting a comparison of the the number of samples $M$ and the the average run time, which attains a similar slope of approximately $-1$.

\section{Conclusion}
\label{sec:conc}

McKean-Vlasov stochastic differential equations (MVSDEs) are an important class of processes, that are used in a range of applications.
A recent study of these processes has been on inference, related to either parameter estimation, or approximations of their corresponding
invariant measure.  The focus of this work was to develop a method that is able to unbiasedly approximate invariant measure of MVSDEs,
which are commonly subject to a bias resulting from a discretization scheme. We consider an Euler-Maruyama discretization and present
an unbiased algorithm motivated from the unbiased Monte Carlo algorithms, that exploit variance reduction techniques. We were firstly able to
demonstrate the ergodicity of various processes we consider to an invariant measure at a geometric rate. To the best of our knowledge these are
the first such results, in the discrete-time setting. From this we proved that our estimator is  unbiased. We presented
various numerical experiments to verify our theory on a range of MVSDEs. These includes a Currie-Weiss model, a 3D neuron model and a parameter 
estimation problem. Our motivation was to consider MVSDEs which omitted an invariant measure, and some where an approximation was required.
\bigskip \\
In terms of future work there are a number of interesting directions one can take. 
\begin{itemize}
\item  A first direction is to consider higher-order discretization schemes,
which have more favourable strong and weak error rates. Examples of this would be splitting order schemes, such as BAOAB and UBU. This has been
explored in the following works \cite{chada,chada2}, which also used to develop unbiased estimator for sampling.  
\item A second direction would be the consideration
of neural MVSDEs \cite{yang}, which are motivated form the recent directions of neural SDEs and neural ODEs, used for approximations within deep-learning.
This is a recent field, with considerable potential in diffusion models. Presenting new schemes at handling various biases would prove useful. 
\item One could consider the application of such unbiased methods to McKean-Vlasov stochastic partial differential equations (MVSPDEs).
This is very much an open direction as very little work has been conducted, both numerically and theoretically.
\item Finally it would be of interest to verify if our unbiased estimator is of finite variance. This computation is not so trivial, and goes beyond the work of this article. We envision the proof procedure would follow similarly to that in \cite{siddig}. \textcolor{black}{This is also related to proving that our unbiased estimator, attains finite computational time. This has been discussed in Section \ref{sec:theo_disc} where we leave this for an additional paper}.
\end{itemize}
\appendix

\section{Ergodicity Results}\label{app:erg}



\subsection{Notation}
For a matrix $A\in\mathbb{R}^{d\times d}$  define the norm $\|A\| = \sup_{x\in\mathbb{R}^d\backslash\{0\}}\frac{|Ax|}{|x|}$ which is equal to the absolute value of the largest eigenvalue of $A$.
Denote by $\mathcal{P}_0(\mathbb{R}^d)$ the set of probability measures on $\mathbb{R}^d$, and for $i\in\mathbb{N}$ define the set of probability measures with finite $i$-th moment as
$\mathcal{P}_i(\mathbb{R}^d) = \left\{ \mu\in \mathcal{P}_0(\mathbb{R}^d) : \int |x|^i\mu(dx) < \infty \right\}$. In order for us to characterize the notion of ergodicity, we require ergodicity with respect to a metric. One metric we will consider for this work is the $i$-Wasserstein distance, which is provided in the following definition. For a random variable $X$ defined on a probability space with probability measure $\mathbb{P}$ we denote by $\mathcal{L}_X=\mathbb{P}\circ X^{-1}$ the law of $X$. Let $\mu$ be any sigma-finite measure on the measurable space $(\mathbb{R}^d,\mathcal{B}(\mathbb{R}^d))$, $\mathcal{B}(\mathbb{R}^d)$ are the Borel sets,  and let $\varphi:\mathbb{R}^d\rightarrow\mathbb{R}$ be $\mu-$integrable, then we write $\mu(\varphi)=\int_{\mathbb{R}^d}\varphi(x)\mu(dx)$.

\begin{defn}[$i$-Wasserstein distance]
\label{def:wass}
The $i$-Wasserstein distance between two measures  $\mu, \nu \in \mathcal{P}_{\textcolor{black}{i}}(\mathbb{R}^d)$ is defined as
$$
\mathcal{W}_i(\mu,\nu) =  \left(\inf_{\gamma\in\Gamma(\mu,\nu)}\int_{\mathbb{R}^{2d}}|x-y|^i\gamma\left(d(x,y)\right)\right)^{1/i},$$
where 
$$
\Gamma(\mu,\nu) = \left\{\gamma\in\mathcal{P}_0(\mathbb{R}^d\times\mathbb{R}^d) : \int_{A\times \mathbb{R}^d} \gamma\left(d(x,y)\right) = \mu(A), \int_{\mathbb{R}^d\times A} \gamma\left(d(x,y)\right) = \nu(A) \right\},
$$
 is the set of couplings of $\mu$ and $\nu$. 
\end{defn}

\begin{defn}[$\mathcal{M}$ distance]
Let $\mu,\nu\in\mathcal{P}_1(\mathbb{R}^d)$, define the $\mathcal{M}$ distance between the two measures  $\mu, \nu$ by
$$\mathcal{M}(\mu,\nu) = \sup_{\|f\|_{\mathrm{Lip}}\leq 1} \left| \int_{\mathbb{R}^d} f(x) \mu(dx) - \int_{\mathbb{R}^d} f(x)\nu(dx) \right|.$$
\end{defn}
We utilize the metric $\mathcal{M}$ and the metrics $\mathcal{W}_i$ in the proofs below. The $\mathcal{M}$ and $\mathcal{W}_1$ metrics are related by the inequality $\mathcal{M}(\mu,\nu)\leq \mathcal{W}_1(\mu,\nu)$ which follow from the Kantorovich duality:
$$
\mathcal{W}_1(\mu,\nu) = \sup_{|f|_{\mathrm{Lip}}\leq 1} \left| \int_{\mathbb{R}^d} f(x) \mu(dx) - \int_{\mathbb{R}^d} f(x)\nu(dx) \right|.
$$
Jensen's inequality guarantees that $\mathcal{W}_1(\mu,\nu)\leq\mathcal{W}_2(\mu,\nu)$. For random variables $X,Y$ defined on the same probability space we have $\mathcal{W}_2(\mathcal{L}_X,\mathcal{L}_Y)\leq \mathbb{E}[|X-Y|^2]^{1/2}$. The spaces $(\mathcal{P}_i(\mathbb{R}^d),\mathcal{W}_i)$ are complete metric spaces. 
 Throughout the proofs below, we use the symbol $C$ for generic constants and its value may change from one line to another.  Dependencies on various model and simulation parameters will be stated as needed.

\subsection{Outline of the Results and Structure}

In the appendix we prove several results associated to various processes.  We recall that the original
continuous-time process $\{X_t\}_{t\geq 0}$ is governed by the dynamics \eqref{eq:sde}.  We will
then denote by $\{\tilde{X}_t\}_{t\in\{0,\Delta_l,\dots}$ as the exact Euler-Maruyama time-discretization as 
featured in \eqref{eq:euler}.  We will also have to analyze the interacting particle system as 
described in Algorithm \ref{alg:basic_method} which is denoted as $\{\check{X}_t^i\}_{(i,t)\in\{1,\dots,N\}\times\{0,\Delta_l,\dots\}}$.  Note that $N$ is fixed here.
Finally we will consider  $\{\overline{X}_t\}_{t\in\{0,\Delta_l,\dots\}}$ which the Euler-Maruyama time-discretization of the SDE \eqref{eq:sde},  except that we plug-in the laws of the SDE approximated by\\
$\{\check{X}_t^i\}_{(i,t)\in\{1,\dots,N\}\times\{0,\Delta_l,\dots\}}$. \textcolor{black}{This is provided exactly in \eqref{eq:X_tilde_N_1}. }

We now prove a series of results which are needed for our main results in Section \ref{sec:theory} in the main text.
We begin with Proposition \ref{prop:wang_hs_verification} which essentially implies that there is unique
stationary distribution of the processes $\{X_t\}_{t\geq 0}$ and is summarized in Theorem \ref{thm:ergodic_continuous}.  We then turn to the exact Euler-Maruyama time-discretization 
$\{\tilde{X}_t\}_{t\in\{0,\Delta_l,\dots}$  for which we prove, in Theorem \ref{thm:ergod_1}, existence of a unique stationary distribution $\pi^l$.  Theorem \ref{eq:invaraint_delta_1} shows that $\pi^l$ converges to $\pi$ as $l-$grows in $2-$Wasserstein distance.   In Theorem \ref{thm:ergodic_2} we give a convergence theorem
for the empirical measures $\mu_t^{l,N}$ associated to the particle system $\{\check{X}_t^i\}_{(i,t)\in\{1,\dots,N\}\times\{0,\Delta_l,\dots\}}$, in terms of the convergence in $N$ and $t$ in expected $2-$Wasserstein distance with $\pi^l$.  Theorem \ref{thm:2nd} shows that $\{\overline{X}_t\}_{t\in\{0,\Delta_l,\dots\}}$ has a unique stationary distribution $\Pi^l$ as $t$ grows and a rather important law of large numbers on a time-discrete grid (used in the proof of Theorem \ref{theo:ub}).  Corollary \ref{cor:main_cor} considers the case that $N=N_l$ and a convergence
in $2-$Wasserstein distance of $\Pi^l$ to $\pi$.  The results should be read in order and proofs of later results rely on earlier  ones.

%
%

\subsection{Ergodicity}
\begin{prop}\label{prop:wang_hs_verification}
    Assume \hyperref[assump:A1]{(A1-2)}. Then there exist constants $C_j<+\infty$, $j\in\{1,\dots,4\}$ with $C_3>C_4$ such that for any $(x,y,\mu,\nu)\in(\mathbb{R}^d)^2\times \mathcal{P}_2(\mathbb{R}^d)^2$
    \begin{equation}\label{eq:wang_h1}
         |b(x,\bar{\xi}_2(x,\mu)) - b(y,\bar{\xi}_2(y,\nu))|\leq C_1|x-y| + C_1\mathcal{M}(\mu,\nu),
    \end{equation}
    \begin{equation}\label{eq:wang_h1_drift}
         |a(x,\bar{\xi}_1(x,\mu)) - a(y,\bar{\xi}_1(y,\nu))|\leq C_2|x-y| + C_2\mathcal{M}(\mu,\nu),
    \end{equation}
    \begin{equation}\label{eq:wang_h2_prime}
        \begin{split}
            &2\langle a(x,\bar{\xi}_1(x,\mu))-a(y,\bar{\xi}_1(y,\nu)), x-y\rangle + |b(x,\bar{\xi}_2(x,\mu)) - b(y,\bar{\xi}_2(y,\nu))|^2\\
            &\leq -C_3|x-y|^2 + C_4\mathcal{M}(\mu,\nu)^2,
        \end{split}
    \end{equation}
    \begin{equation}\label{eq:wang_h3}
        \sup_{\mu\in \mathcal{P}_2(\mathbb{R}^d)}|a(0,\bar{\xi}_1(0,\mu))|< \infty , \sup_{\mu\in \mathcal{P}_2(\mathbb{R}^d)}|b(0,\bar{\xi}_2(0,\mu))| < \infty.
    \end{equation}
\end{prop}
\begin{proof}
For the first inequality, using the assumption that $b$ is Lipschitz and the triangular inequality we have
    \begin{align*}
            |b(x,\bar{\xi}_2(x,\mu)) - b(y,\bar{\xi}_2(y,\nu))|&\leq |b|_{\mathrm{Lip}} |x-y| + |b|_{\mathrm{Lip}}\left|\int_{\mathbb{R}^d} \xi_2(x,z)\mu(dz)  - \int_{\mathbb{R}^d} \xi_2(y,z)\mu(dz)\right| \\&+ |b|_{\mathrm{Lip}}\left|\int_{\mathbb{R}^d} \xi_2(y,z)\mu(dz)  - \int_{\mathbb{R}^d} \xi_2(y,z)\nu(dz)\right|.
    \end{align*}
    The second term in the line is bounded by $|\xi_2|_{\mathrm{Lip}}|x-y|$ and the third term is bounded by $\|\xi_2\|_{\mathrm{Lip}}\mathcal{M}(\mu,\nu)$. Thus inequality  \eqref{eq:wang_h1} holds with $C_1=|b|_{\mathrm{Lip}}(1+\|\xi_2\|_{\mathrm{Lip}})$. Inequality \eqref{eq:wang_h1_drift} follows analogously. For \eqref{eq:wang_h2_prime} we write
    \begin{equation*}
    \begin{split}
        &\langle a(x,\bar{\xi}_1(x,\mu))-a(y,\bar{\xi}_1(y,\nu)), x-y\rangle \\=& \langle a(x,\bar{\xi}_1(x,\mu))-a(y,\bar{\xi}_1(x,\mu)), x-y\rangle + \langle a(y,\bar{\xi}_1(x,\mu))-a(y,\bar{\xi}_1(y,\nu)), x-y\rangle.
    \end{split}
    \end{equation*}
    Let $A=\sup_{x\in\mathbb{R}^{d+1}}\sup_{|y|=1} y^{\top}\nabla_1 a(x)y$ and $B=\|\nabla_2 a\|$. The first term is bounded above by $A|x-y|^2$. For the second term, we use the Cauchy inequality and calculations similar to the one used to prove the first inequality
    \begin{equation*}
        \begin{split}
            \langle a(y,\bar{\xi}_1(x,\mu))-a(y,\bar{\xi}_1(y,\nu)), x-y\rangle
            \leq& |x-y|\left|a(y,\bar{\xi}_1(x,\mu))-a(y,\bar{\xi}_1(y,\nu))\right|\\
            \leq& B\|\xi_1\|_{\mathrm{Lip}}|x-y|\left(|x-y|+ \mathcal{M}(\mu,\nu)\right)\\
            \leq& \frac{3}{2}B\|\xi_1\|_{\mathrm{Lip}}|x-y|^2 + \frac{1}{2}B\|\xi_1\|_{\mathrm{Lip}}\mathcal{M}(\mu,\nu)^2.
        \end{split}
    \end{equation*}
    Therefore
    \begin{equation*}
        \begin{split}
            &2\langle a(x,\bar{\xi}_1(x,\mu))-a(y,\bar{\xi}_1(y,\nu)), x-y\rangle + |b(x,\bar{\xi}_2(x,\mu)) - b(y,\bar{\xi}_2(y,\nu))|^2\\
            \leq& (2A+3B\|\xi_1\|_{\mathrm{Lip}}+2C_1^2)|x-y|^2 + (B\|\xi_1\|_{\mathrm{Lip}}+2C_1^2)\mathcal{M}(\mu,\nu)^2,
        \end{split}
    \end{equation*}
    and \eqref{eq:wang_h2_prime} follows by Assumption \hyperref[assump:A2]{(A2)}. Inequality \eqref{eq:wang_h3} follows from the continuity of the function $a$ and the boundedness of the function $\bar{\xi}_1$
    $$\sup_{\mu\in\mathcal{P}_2(\mathbb{R}^d)}a(0,\bar{\xi}_1(0,\mu))\leq \sup_{|x|\leq \|\xi_1\|}a(0,x)<\infty.$$
\end{proof}

Proposition \ref{prop:wang_hs_verification} verifies the conditions needed for \cite[Theorem 3.1]{wang}, which we now state.

\begin{theorem}\label{thm:ergodic_continuous}
    Assume \hyperref[assump:A1]{(A1-2)}. Then there exists a unique $\pi\in\mathcal{P}_2(\mathbb{R}^d)$
such that
$$
\lim_{t\to\infty}\mathcal{W}_2(\mu_t,\pi)=0.
$$
\end{theorem}
Let $\Delta_l>0$, unless explicitly stated the processes below in this section depend implicitly on $\Delta_l$. Define $\Delta_l W_t= W_{t+\textcolor{black}{\Delta_l}}-W_t$. 
Recall the discrete process $\{\widetilde{X}_t\}_{t\in\{0,\Delta_l,\dots\}}$ 
$$
\tilde{X}_{t+\Delta_l} = \tilde{X}_t + a(\tilde{X}_t,\bar{\xi}_1(\tilde{X}_t,\mu_t^l))\Delta + b(\tilde{X}_t,\bar{\xi}_1(\tilde{X}_t,\mu_t^l))\Delta W_{t},
$$
where $\mu_t^l$ is the law of $\tilde{X}_t$.  We follow the proof of Theorem 3.1 in \cite{wang} but in a discrete manner to show that the process $\{\tilde{X}_t\}_{t\in\{0,\Delta_l,\dots\}}$ has a unique invariant measure.  Define 
$$
\Delta^{\star} = \min\left\{\frac{C_3-C_4}{2C_2} , \frac{C_3}{4C_2^2}\right\}.
$$
$\Delta^{\star}$ will serve as a threshold for the values of $\Delta_l$ for which the following proofs will be valid.


\begin{theorem}\label{thm:ergod_1}
    Assume \hyperref[assump:A1]{(A1-2)} and that $\Delta_l<\Delta^{\star}$. Then there exists a unique
$\pi^l\in\mathcal{P}_2(\mathbb{R}^d)$ such that
$$
\lim_{t\to\infty}\mathcal{W}_2(\mu_t^l,\pi^l)=0.
$$
Moreover,  $\pi^l$ is independent of $\mu_0^l$ and if $\mu_0^l=\pi^l$ then $\tilde{X}_t\sim\pi^l$ for every $t\in\{0,\Delta_l,\dots\}$.
\end{theorem}
\begin{proof}
    Let $s\in \{\Delta_l,2\Delta_l,\dots\}$. Let $\{\tilde{Y}_t\}_{t\in\{0,\Delta_l,\dots\}}$ be the discrete process defined by $\tilde{Y}_0\sim \mu_s^l$ and 
    \begin{align*}
\tilde{Y}_{t+\Delta_l}&= \tilde{Y}_t + a(\tilde{Y}_t,\bar{\xi}_1(\tilde{Y}_t,\nu_t^l))\Delta_l + b(\tilde{Y}_t,\bar{\xi}_1(\tilde{Y}_t,\nu_t^l))\Delta_l W_{t},
    \end{align*}
where we remark that \textcolor{black}{ coupling between $\tilde{X}_0$ and $\tilde{Y}_0$
 is chosen in such a way that this holds.} 
$\nu_t^l$ is the law of $\tilde{Y}_t$.  As $\tilde{Y}_t$ is discrete and follows the same iteration as $\tilde{X}_t$ we have $\nu_t^l = \mu_{s+t}^l$. For every $t\in\{0,\Delta_l,\dots\}$ we have
    \begin{equation*}
        \begin{split}
            \mathbb{E}[|\tilde{X}_{t+\Delta} - \tilde{Y}_{t+\Delta}|^2]
            &= \mathbb{E}[ |\tilde{X}_t - \tilde{Y}_t|^2] + \mathbb{E}[|a(\tilde{X}_t,\bar{\xi}_1(\tilde{X}_t,\mu_t^l)) - a(\tilde{Y}_t,\bar{\xi}_1(\tilde{Y}_t,\nu_t^l))|^2]\Delta_l^2\\
            &+\mathbb{E}[2\langle a(\tilde{X}_t,\bar{\xi}_1(\tilde{X}_t,\mu_t^l)) - a(\tilde{Y}_t,\bar{\xi}_1(\tilde{Y}_t,\nu_t^l)),\tilde{X}_t-\tilde{Y}_t\rangle \\&+ |b(\tilde{X}_t,\bar{\xi}_2(\tilde{X}_t,\mu_t^l)) - b(\tilde{Y}_t,\bar{\xi}_2(\tilde{Y}_t,\nu_t^l))|^2]\Delta_l\\
            &\leq  (1-C_3\Delta_l + C_2\Delta_l^2)\mathbb{E}[ |\tilde{X}_t - \tilde{Y}_t|^2] + (C_4\Delta_l + C_2\Delta_l^2)\mathcal{W}_2(\mu_t^l,\nu_t^l)^2.\\
            &\leq  (1-C_3\Delta_l + C_2\Delta_l^2 + C_4\Delta_l + C_2\Delta_l^2)\mathbb{E}[ |\tilde{X}_t - \tilde{Y}_t|^2],
        \end{split}
    \end{equation*}
    where the constants $C_2,C_3,C_4$ are as in Proposition \ref{prop:wang_hs_verification}.  As $\Delta_l<\Delta^{\star}$ we have
$$
\epsilon=C_3 - C_2\Delta_l - C_4 - C_2\Delta_l > 0.
$$
   Therefore for every $t\in\{0,\Delta_l,\dots\}$   
\begin{equation}\label{eq:w_inq_2}
   \begin{split}
       \mathbb{E}[|\tilde{X}_{t} - \tilde{Y}_{t}|^2] = &\mathbb{E}[|\tilde{X}_{0} - \tilde{Y}_{0}|^2]\prod_{k=0}^{t/\Delta_l-1}\frac{\mathbb{E}[|\tilde{X}_{(k+1)\Delta_l} - \tilde{Y}_{(k+1)\Delta_l}|^2]}{\mathbb{E}[|\tilde{X}_{k\Delta_l} - \tilde{Y}_{k\Delta_l}|^2]}\\
       \leq& \mathcal{W}_2(\mu_0^l,\nu_0^l)^2 (1-\epsilon\Delta_l)^{t/\Delta_l}\\
       \leq &\mathcal{W}_2(\mu_0^l,\nu_0^l)^2e^{-\epsilon t}.
   \end{split}
   \end{equation}
   This implies that
   \begin{equation}\label{eq:w_inq_1}
       \mathcal{W}_2(\mu_t^l,\mu_{t+s}^l)^2\leq \mathcal{W}_2(\mu_0^l,\mu_s^l)^2e^{-\epsilon t}\leq 4\sup_{s\in
\{0,\Delta_l,\dots\}}\mathbb{E}[|\tilde{X}_s|^2]e^{-\epsilon t}.
   \end{equation}
    To bound $\mathbb{E}[|\tilde{X}_t|^2]$ we follows similar calculations as follows
    \begin{equation}\label{eq:X_tilde_bound_1}
        \begin{split}
            \mathbb{E}[|\tilde{X}_{t+\Delta_l}|^2]=&\mathbb{E}[|\tilde{X}_t|^2] + \Delta_l\mathbb{E}[2\langle a(\tilde{X}_t,\bar{\xi}_1(\tilde{X}_t,\mu_t^l)),\tilde{X}_t\rangle + |b(\tilde{X}_t,\bar{\xi}_1(\tilde{X}_t,\mu_t^l))|^2]\\
            =& \mathbb{E}[|\tilde{X}_t|^2] + \Delta_l\mathbb{E}[2\langle a(\tilde{X}_t,\bar{\xi}_1(\tilde{X}_t,\mu_t^l)) - a(0,\bar{\xi}_1(0,\mu_t^l)),\tilde{X}_t\rangle \\&+ |b(\tilde{X}_t,\bar{\xi}_1(\tilde{X}_t,\mu_t^l))-b(0,\bar{\xi}_1(0,\mu_t^l))|^2] \\
            +& \Delta_l\mathbb{E}[2\langle b(\tilde{X}_t,\bar{\xi}_1(\tilde{X}_t,\mu_t^l)),b(0,\bar{\xi}_1(0,\mu_t^l)) \rangle - |b(0,\bar{\xi}_1(0,\mu_t^l))|^2] \\&+ \Delta_l\mathbb{E}[\langle a(0,\bar{\xi}_1(0,\mu_t^l)),\tilde{X}_t\rangle]\\
            \leq&(1-C_3\Delta_l)\mathbb{E}[|\tilde{X}_t|^2] + C\Delta_l\mathbb{E}[|\tilde{X}_t|] + C\Delta_l\\
            \leq &(1-C_3\Delta_l/2)\mathbb{E}[|\tilde{X}_t|^2] + C\Delta_l,
        \end{split}
    \end{equation}
    where we used the boundedness of $a(0,\bar{\xi}_1(0,\mu_t^l))$ and $b(0,\bar{\xi}_1(0,\mu_t^l))$,  the Cauchy inequality, and the inequality $$x\leq \frac{C}{2C_3} + \frac{C_3}{2C}x^2.$$ 
    Iterating inequality \eqref{eq:X_tilde_bound_1} yields
    \begin{equation}\label{eq:X_inq_1}
        \mathbb{E}[|\tilde{X}_t|^2]\leq (1-C_3\Delta_l/2)^{t/\Delta_l}\mathbb{E}[|\tilde{X}_0|^2] + C\Delta_l\sum_{k=0}^{\infty} (1-C_3\Delta_l/2)^k\leq e^{-C_3t/2}\mathbb{E}[|\tilde{X}_0|^2] + \frac{2C}{C_3}.
    \end{equation}
     Therefore the sequence $\{\mu_t^l\}_{t\in\{0,\Delta_l,\dots\}}$ is a Cauchy sequence on the complete metric space $(\mathcal{P}_2,\mathcal{W}_2)$, thus there exists an invariant measure $\pi^l$ that satisfies $\lim_{t\to\infty}\mathcal{W}_2(\mu_t^l,\pi^l)$=0. Taking $s\to\infty$ in the first half of inequality \eqref{eq:w_inq_1} shows that if $\mu_0^l=\pi^l$ then $\tilde{X}_t\sim\pi^l$. 
     \par
     To prove uniqueness and independence of the initial distribution of the process, let $\tilde{\nu}$ be an invariant measure and define the processes $\tilde{Y}_t$ as above but with $\tilde{Y}_0\sim \tilde{\nu}$ and suppose that $\mu_0^l=\pi^l$. By invariance of $\pi^l$ we have $\tilde{Y}_t\sim\tilde{\nu}$ for all $t\in\{0,\delta_l,\dots\}$. From inequality \eqref{eq:w_inq_2} we have
   $$\mathcal{W}_2(\pi^l,\tilde{\nu})^2\leq \mathcal{W}_2(\pi^l,\tilde{\nu})^2e^{-\epsilon t},$$
   and the claim follows by taking $t\to\infty$. 
\end{proof}

\begin{cor}
Assume \hyperref[assump:A1]{(A1-2)}.  Let $\overline{\Delta}<\Delta^{\star}$ and $\Delta_l<\overline{\Delta}$. 
The bounds in \eqref{eq:w_inq_2},\eqref{eq:w_inq_1}, and \eqref{eq:X_inq_1} depend only on $\overline{\Delta}$, $\mathbb{E}[|\tilde{X_0}|^2]$, and the norms of the coefficients $a,b,\xi_1,\xi_2$.
\end{cor}

\begin{theorem}\label{eq:invaraint_delta_1}
Assume \hyperref[assump:A1]{(A1-2)}.  Then we have
    $$
\lim_{l\rightarrow\infty}\mathcal{W}_2(\pi^l,\pi)=0.
$$
\end{theorem}
\begin{proof}
    Without loss of generality assume $\Delta_l<\Delta^{\star}$.
    First, by \cite[Theorem 3.1]{wang} we have $\sup_{s\geq0}\mathbb{E}[|X_s|^2]<\infty$. Second, by using the inquality $|x+y|^2\leq 2|x|^2+2|y|^2$, Lipschitz properties of the functions $a$ and $b$, Proposition \ref{prop:wang_hs_verification}, Cauchy inequality, Ito isometry, and Fubini's Theorem, we have that for $t>s>0$:
    \begin{equation}\label{eq:X_X_ilde_1}
        \mathbb{E}[|X_t-X_s|^2] \leq C\mathbb{E}\left[\left|\int_s^t(|X_u|+1)du\right|^2\right] + C\mathbb{E}\left[\int_s^t(|X_u|+1)^2du\right]<C(t-s),
    \end{equation}
    with $C$ independent of $t$ and $s$. Write
    \begin{equation*}
        X_{t+\Delta_l} - \tilde{X}_{t+\Delta_l} = A + B,
    \end{equation*}
    where
       \begin{align*}
    A:&=X_t - \tilde{X}_t + (a(X_t,\bar{\xi}_1(X_t,\mu_t))-a(\tilde{X}_t,\bar{\xi}_1(\tilde{X}_t,\mu_t^l)))\Delta_l \\&+ (b(X_t,\bar{\xi}_1(X_t,\mu_t))-b(\tilde{X}_t,\bar{\xi}_2(\tilde{X}_t,\mu_t^l)))\Delta_l W_t,
        \end{align*}
    and 
    \begin{align*}B:&=\int_t^{t+\Delta_l} (a(X_u,\bar{\xi}_1(X_u,\mu_u)) - a(X_t,\bar{\xi}_1(X_t,\mu_t)))du \\&+ \int_t^{t+\Delta_l} (b(X_u,\bar{\xi}_2(X_u,\mu_u)) - b(X_t,\bar{\xi}_2(X_t,\mu_t)))dW_u.
    \end{align*}

    Following the proof of Theorem \ref{thm:ergod_1} there exists $\epsilon>0$ independent of $\Delta_l$ that satisfies
    $$\mathbb{E}[|A|^2]\leq (1-\epsilon\Delta_l)\mathbb{E}[|X_t-\tilde{X}_t|^2].$$
    Using \eqref{eq:X_X_ilde_1} and Ito isometry we have
    $$
\mathbb{E}[|B|^2] \leq C\Delta_l \sup_{0\leq s\leq \Delta_l}\mathbb{E}[|X_{t+s}-X_t|^2]\leq C\Delta_l^2. 
$$
    Using \eqref{eq:X_X_ilde_1} and It\^{o} isometry we have that $\mathbb{E}[\langle A,B\rangle]$ is equal to
    \begin{equation*}
        \begin{split}
            &\mathbb{E}\Bigg[\left((b(X_t,\bar{\xi}_1(X_t,\mu_t))-b(\tilde{X}_t,\bar{\xi}_2(\tilde{X}_t,\mu_t^l)))\Delta_l W_t\right)^{\top} \\&\times \int_t^{t+\Delta_l} (b(X_u,\bar{\xi}_2(X_u,\mu_u)) - b(X_t,\bar{\xi}_2(X_t,\mu_t)))dW_u\Bigg]\\
            \leq& \mathbb{E}\left[\left|(b(X_t,\bar{\xi}_1(X_t,\mu_t))-b(\tilde{X}_t,\bar{\xi}_2(\tilde{X}_t,\mu_t^l)))\right|\int_t^{t+\Delta_l}\left|b(X_u,\bar{\xi}_2(X_u,\mu_u)) - b(X_t,\bar{\xi}_2(X_t,\mu_t))\right|du \right]\\
            \leq& C\mathbb{E}\left[ \left(|X_t-\tilde{X}_t| + \mathcal{W}_1(\mu_t,\mu_t^l)\right)\int_t^{t+\Delta_l} \left(|X_u-X_t| + \mathcal{W}_1(\mu_u,\mu_t)\right)du \right]\\
            \leq &C\Delta_l\mathbb{E}[|X_t-\tilde{X}_t|^2]^{1/2}\sup_{0\leq s\leq \Delta_l}\mathbb{E}[|X_{t+s}-X_t|^2]^{1/2}\\
            \leq & C\Delta_l^{3/2}\mathbb{E}[|X_t-\tilde{X}_t|^2]^{1/2}\\
            \leq& \frac{\epsilon\Delta_l}{2}\mathbb{E}[|X_t-\tilde{X}_t|^2] + C\frac{\Delta_l^2}{2\epsilon}.
        \end{split}
    \end{equation*}
    To deduce the last three lines we used the inequalities 
   $$ 
   \mathcal{W}_1(\mu_t,\mu_t^l)\leq\mathbb{E}[|X_t-\tilde{X}_t|]^{1/2}, \quad \mathcal{W}_1(\mu_u,\mu_t)\leq\mathbb{E}[|X_u-X_t|]^{1/2}, \quad x\leq \frac{\epsilon}{2C\sqrt{\Delta_l}}x^2 + \frac{C\sqrt{\Delta_l}}{2\epsilon}.
   $$
    Therefore
    \begin{equation*}
        \begin{split}
            \mathbb{E}[|X_{t+\Delta_l}-\tilde{X}_{t+\Delta_l}|^2]\leq& \mathbb{E}[|A|^2] + 2\mathbb{E}[\langle A,B\rangle] + \mathbb{E}[|B|^2]\\
            \leq& (1-\epsilon\Delta_l/2)\mathbb{E}[|X_t-\tilde{X}_t|^2] + C\Delta_l^2.
        \end{split}
    \end{equation*}
   Iterating this last inequality yields
   $$
\mathbb{E}[|X_t-\tilde{X}_t|^2]\leq (1-\epsilon\Delta_l/2)^{t/\Delta_l}\mathbb{E}[|X_0-\tilde{X}_0|^2] + C\Delta_l^2\sum_{k=0}^{\infty}(1-\epsilon\Delta_l/2)^k\leq Ce^{-\epsilon t/2} + C\Delta_l.
$$
    Using $\mathcal{W}_2(\mu_t,\mu_t^l)\leq\mathbb{E}[|X_t-\tilde{X}_t|^2]$ and taking $t\to\infty$ then $l\rightarrow\infty$ proves the claim.
\end{proof}


Consider $(i,k)\in\{1,\dots,N\}\times\mathbb{N}_0$
\begin{equation}\label{eq:ips}
\check{X}^{i}_{(k+1)\Delta_l} = \check{X}^{i}_{k\Delta_l} + a(\check{X}^{i}_{k\Delta_l},\bar{\xi}_1(\check{X}^{i}_{k\Delta_l},\mu_{k\Delta_l}^{l,N}))\Delta_l + b(\check{X}^{i}_{k\Delta_l},\bar{\xi}_1(\check{X}^{i}_t,\mu^{l,N}_{k\Delta_l}))\Delta_l W^i_{k\Delta_l},
\end{equation}
where $\mu^{l,N}_{k\Delta_l} = \frac{1}{N}\sum_{j=1}^N\delta_{\check{X}^{i}_{k\Delta_l}}$, $\check{X}^{i}_0=x_0$, $i\in\{1,\dots,N\}$ and $\{W^i_{k\Delta_l}\}_{i\in\{1,\dots,N\}}$ are independent standard Brownian motions. 

\begin{theorem}\label{thm:ergodic_2}
Assume \hyperref[assump:A1]{(A1-A2)} and $\Delta_l<\Delta^{\star}$. Then we have
    $$
\lim_{\substack{N\to\infty\\t\to\infty}}\mathbb{E}[\mathcal{W}_2(\mu^{l,N}_t,\pi^l)^2]=0.
$$
\end{theorem}

\begin{proof}
    Consider the system
    \begin{equation*}
            \tilde{X}^{i}_{t+\Delta_l} = \tilde{X}^{i}_t + a(\tilde{X}^{i}_t,\bar{\xi}_1(\tilde{X}^{i}_t,\mu_t^l))\Delta_l + b(\tilde{X}^{i}_t,\bar{\xi}_1(\tilde{X}^{i}_t,\mu_t^l))\Delta_l W^i_{t},
    \end{equation*}
    with $\tilde{X}_0^i\stackrel{\textrm{ind}}{\sim}\mu_0^l$ for $i\in\{1,\dots,N\}$.
Following the calculations in Theorem \ref{thm:ergod_1} we can show that 
    $$
    \sup_{t\geq0}\mathbb{E}[|\check{X}^{i}_t|^2]<\infty, 
   \quad \sup_{t\geq0}\mathbb{E}[|\tilde{X}^{i}_t|^2]<\infty,
   $$ 
   and
    \begin{equation*}
        \begin{split}
            \mathbb{E}[|\check{X}^{i}_{t+\Delta_l} - \tilde{X}^i_{t+\Delta_l}|^2]
            &= \mathbb{E}[|\check{X}^{i}_t - \tilde{X}^i_t|^2] + \mathbb{E}[|a(\check{X}^{i}_t,\bar{\xi}_1(
\check{X}^{i}_t,\mu_t^{l,N})) - a(\tilde{X}^i_t,\bar{\xi}_1(\tilde{X}^i_t,\mu_t^l))|^2]\Delta_l^2\\
            &+\mathbb{E}[2\langle a(\check{X}^{i}_t,\bar{\xi}_1(\check{X}^{i}_t,\mu_t^{l,N})) - a(\tilde{X}^i_t,\bar{\xi}_1(\tilde{X}^i_t,\mu_t^l)),\check{X}^{i}_t-\tilde{X}^i_t\rangle \\&+ |b(\check{X}^{i}_t,\bar{\xi}_2(\check{X}^{i}_t,\mu_t^{l,N})) - b(\tilde{X}^i_t,\bar{\xi}_2(\tilde{X}^i_t,\mu_t^l))|^2]\Delta_l\\
            &\leq  (1-C_3\Delta_l + C_2\Delta_l^2)\mathbb{E}[ |\check{X}^{i}_t - \tilde{X}^i_t|^2] + (C_4\Delta_l + C_2\Delta_l^2)\mathbb{E}[\mathcal{M}(\mu^{l,N}_t,\mu_t^l)^2].\\
            &\leq  (1-C_3\Delta_l + C_2\Delta_l^2 + C_4\Delta_l + C_2\Delta_l^2)\mathbb{E}[ |\check{X}^{i}_t - \tilde{X}^i_t|^2].
        \end{split}
    \end{equation*}    
    Let $\kappa>0$ such that $$C_3-C_2\Delta_l-(1+\kappa)(C_4+C_2\Delta_l)>0,$$ such a $\kappa$ exists because $\Delta_l<\Delta^{\star}$. Using the following inequality 
    $$(x+y)^2\leq (1+\kappa)x^2+\Big(1+\frac{1}{\kappa}\Big)y^2,$$ 
    we have that for any $f\in\mathcal{C}^{\mathrm{Lip}}(\mathbb{R}^d,\mathbb{R})\cap\mathcal{C}_b(\mathbb{R}^d,\mathbb{R})$
    \begin{equation*}
    \begin{split}
        \mathbb{E}[|\mu^{l,N}_t(f) - \mu_t^l(f)|^2]\leq& (1+\kappa)\mathbb{E}\left[\left|\frac{1}{N}\sum_{i=1}^N\left\{f(\check{X}^{i}_t) - f(\tilde{X}^{i}_t)\right\}\right|^2\right] \\&+ \Big(1+\frac{1}{\kappa}\Big)\mathbb{E}\left[\left|\frac{1}{N}\sum_{i=1}^N\left\{f(\check{X}^{i}_t) - \mathbb{E}[f(\tilde{X}^{i}_t)]\right\}\right|^2\right]\\
        \leq& (1+\kappa)\|f\|_{\mathrm{Lip}}^2\sup_i\mathbb{E}[|\check{X}^{i}_t-\tilde{X}^{i}_t|^2] + \Big(1+\frac{1}{\kappa}\Big)\frac{1}{N}\mathbb{E}\left[(f(\check{X}^{1}_t) - \mathbb{E}[f(\tilde{X}^{1}_t)])^2\right]\\
        \leq& (1+\kappa)\|f\|_{\mathrm{Lip}}^2\sup_i\mathbb{E}[|\check{X}^{i}_t-\tilde{X}^{i}_t|^2] + 4\Big(1+\frac{1}{\kappa}\Big)\|f\|^2\frac{1}{N}.
    \end{split}
    \end{equation*}
In the above calculation,  for the first term after the first inequality we used the inequality 
$$
\left(\frac{1}{N}\sum_{i=1}^Nx_i\right)^2\leq \frac{1}{N}\sum_{i=1}^Nx_i^2,
$$ 
and for the second term we use the fact that the random variables $\tilde{X}^i_t$ are i.i.d.. Now we define a sequence of bounded Lipschitz random functions $f_n$ that satisfy the following inequalities
$$    
\mathcal{M}(\mu^{l,N}_t,\mu_t^l) - \frac{1}{n}\leq \mu^{l,N}_t(f_n) - \mu_t^l(f_n)\leq \mathcal{M}(\mu^{l,N}_t,\mu_t^l), \quad \|f_n\|_{\mathrm{Lip}}\leq 1.
$$
 Since $\mathcal{M}(\mu_t^{l,N},\mu_t^l)\leq2$ we have by dominated convergence 
$$
\mathbb{E}[\mathcal{M}(\mu_t^{l,N},\mu_t^l)^2]= \lim_{n\to\infty}\mathbb{E}[|\mu^{l,N}_t(f_n) - \mu_t^l(f_n)|^2]\leq (1+\kappa)\sup_i\mathbb{E}[|\check{X}^{i}_t-\tilde{X}^{i}_t|^2] + 4\Big(1+\frac{1}{\kappa}\Big)\frac{1}{N}.
$$
    
    Letting $\epsilon = C_3 - C_2\Delta_l - (1+\kappa)(C_4 + C_2\Delta_l)$, we have
    $$
\sup_i\mathbb{E}[|\check{X}^{i}_{t+\Delta} - \tilde{X}^i_{t+\Delta}|^2]\leq (1-\epsilon\Delta)\sup_i\mathbb{E}[ |\check{X}^{i}_t - \tilde{X}^i_t|^2] + \frac{C\Delta_l}{N}.
$$
    Consequently
    \begin{equation*}
        \begin{split}
            \sup_i\mathbb{E}[|\check{X}^{i}_{t} - \tilde{X}^i_{t}|^2]\leq& (1-\epsilon\Delta)^{t/\Delta_l}\sup_i\mathbb{E}[|\check{X}^{i}_{0} - \tilde{X}^i_{0}|^2] + \frac{C\Delta_l}{N}\sum_{k=0}^{\infty}(1-\epsilon\Delta_l)^k \\<& e^{-\epsilon t}\sup_i\mathbb{E}[|\check{X}^{i}_{0} - \tilde{X}^i_{0}|^2] + \frac{C}{\epsilon N}.
        \end{split}
    \end{equation*} 
    Noticing that $\mathcal{W}_2(\mu_t^{l,N},\mu_t^l)^2\leq \sup_i\mathbb{E}[|\check{X}^{i}_{t} - \tilde{X}^i_{t}|^2]$ proves that $$\lim_{\substack{N\to\infty\\t\to\infty}}\mathbb{E}[\mathcal{W}_2(\mu_t^{l,N},\mu_t^l)^2]=0.$$
    Finally, the Theorem statement follows from the inequality 
$$
\mathcal{W}_2(\mu_t^{l,N},\pi^l)^2 \leq 2\mathcal{W}_2(\mu_t^{l,N},\mu_t^l)^2 + 2\mathcal{W}_2(\mu_t^l,\pi^l)^2,
$$ 
and using Theorem \ref{thm:ergod_1}.
\end{proof}

For $k\in\mathbb{N}_0$ set
\begin{equation}\label{eq:X_tilde_N_1}
    \overline{X}_{(k+1)\Delta_l} = \overline{X}_{k\Delta_l} + a(\overline{X}_{k\Delta_l},\bar{\xi}_1(\overline{X}_{k\Delta_l},\mu^{l,N}_{k\Delta_l}))\Delta_l + b(\overline{X}_{k\Delta_l},\bar{\xi}_1(\overline{X}_{k\Delta_l},\mu^{l,N}_{k\Delta_l}))\Delta_l B_{k\Delta_l},
\end{equation}
where the empirical measures have been plugged in from the system \eqref{eq:ips},  $\overline{X}_0=x_0$
 and $B_{k\Delta_l}$ is a standard Brownian motion independent of all random variables.
By conditioning on $\mathscr{L}$ we can follow the same strategy of Theorem \ref{thm:ergod_1} and show that there exists a unique invariant (random) measure $\Pi^l$ for the process defined in \eqref{eq:X_tilde_N_1}. Furthermore, we have
\begin{equation}\label{eq:X_tilde_N_inq_1}
    \mathcal{W}_2(\Pi^l,\mathcal{L}_{\overline{X}_t})^2\leq \mathcal{W}_2(\Pi^l,\mathcal{L}_{\overline{X}_0})^2e^{-\epsilon t}.
\end{equation}
with $\epsilon$ a constant independent of $\mathscr{L}$, which implies $
    \mathcal{W}_2(\Pi^l,\mathcal{L}_{\overline{X}_t})\to0$ both a.s.~and in $\mathbb{L}_2$ as $t\to\infty$. Next, we show that a law of large numbers holds.

\begin{theorem}
\label{thm:2nd}
Assume \hyperref[assump:A1]{(A1-2)} and  $\Delta_l<\Delta^{\star}$. Then there exists a unique $\Pi^l\in\mathcal{P}_2(\mathbb{R}^d)$ such that 
$$
  \mathcal{W}_2(\Pi^l,\mathcal{L}_{\overline{X}_t}) \xrightarrow[t\to\infty]{\mathrm{a.s. \ and \ }\mathbb{L}_2} 0
$$
and
\begin{equation}\label{eq:Pitopi}
\mathcal{W}_2(\Pi^l,\pi^l)\xrightarrow[N\to\infty]{\mathbb{L}_2}0.
\end{equation}
In addition, for any $\varphi\in\mathcal{C}^{\mathrm{Lip}}(\mathbb{R}^d,\mathbb{R})$
 \begin{equation}\label{eq:lln}
\mathbb{E}\left[\frac{1}{I}\sum_{t=1}^I \varphi(\overline{X}_t)\bigg|\mathscr{L}\right] \xrightarrow[I\to\infty]{\mathrm{a.s. \ and \ }\mathbb{L}_2} \Pi^l(\varphi).
\end{equation}
\end{theorem}
\begin{proof}
    The existence and uniqueness of $\Pi^l$ is established above and we defer the proof of \eqref{eq:Pitopi} at the end.
For $\varphi\in\mathcal{C}^{\mathrm{Lip}}(\mathbb{R}^d,\mathbb{R})$ we have almost surely
   \begin{align*}
\left| 
\mathbb{E}\left[\frac{1}{I}\sum_{t=1}^I \varphi(\overline{X}_t)\bigg|\mathscr{L}\right]-\Pi^l(\varphi) \right|&\leq 
\frac{1}{I}\sum_{t=1}^I\left|
\mathbb{E}\left[\frac{1}{I}\sum_{t=1}^I \varphi(\overline{X}_t)\bigg|\mathscr{L}\right]
-\Pi^l(\varphi)
 \right|
 \\
 &\leq\frac{|\varphi|_{\mathrm{Lip}}}{I}\sum_{t=1}^M\mathcal{W}_2(\mathcal{L}_{\overline{X}_t},\Pi^l),
\end{align*}
and one can conclude \eqref{eq:lln} by Cesaro averages.

For  \eqref{eq:Pitopi},  recall the discrete-time process in \eqref{eq:X_tilde_N_1} and define the process
$k\in\mathbb{N}_0$
 $$
    \overline{Z}_{(k+1)\Delta_l} = \overline{Z}_{k\Delta_l} + a(\overline{X}_{k\Delta_l},\bar{\xi}_1(\overline{Z}_{k\Delta_l},\mu^{l}_{k\Delta_l}))\Delta_l + b(\overline{Z}_{k\Delta_l},\bar{\xi}_1(\overline{Z}_{k\Delta_l},\mu^{l}_{k\Delta_l}))\Delta_l B_{k\Delta_l},
$$
$\overline{Z}_0=x_0$.
    Following the calculations of Theorem \ref{thm:ergod_1} we have that $$\sup_{N\in\mathbb{N}}
\sup_{s\in\{0,\Delta_l,\dots\}}\mathbb{E}[|\overline{X}_s|^2]<\infty, \quad \sup_{N\in\mathbb{N}}\sup_{s\in\{0,\Delta_l,\dots\}}\mathbb{E}[|\overline{Z}_s|^2]<\infty,
$$ 
and
    \begin{equation}\label{eq:X_tilde_N_1_2}
    \begin{split}
        \mathbb{E}[|\overline{X}_{t+\Delta_l} - \overline{Z}_{t+\Delta_l}|^2]
        \leq (1-C_3\Delta_l + C_2\Delta_l^2)\mathbb{E}[ |\overline{X}_t - \overline{Z}_t|^2] + (C_4\Delta_l + C_2\Delta_l^2)\mathbb{E}[\mathcal{M}(\mu^{l,N}_t,\mu_t^l)^2],
    \end{split}
    \end{equation}
    Let $\zeta>0$  by Theorem \ref{thm:ergodic_2} there exist  $s_1\in\{0,\Delta_l,\dots\}$ and $A\in\mathbb{N}$ such that 
$$
\mathbb{E}[\mathcal{M}(\mu^{l,N}_t,\mu^l_t)^2]\leq \mathbb{E}[\mathcal{W}_2(\mu^{l,N}_t,\mu^l_t)^2] <\zeta,
$$ 
for all $t>s_1$ and $N>A$. Let $s_2\in\{0,\Delta_l,\dots\}$ satisfy $e^{(-C_3+C_2\Delta_l)s_2}<\zeta$ and let $s=\max(s_1,s_2)$. For every $N>A$ and $t>2s$ the inequality \eqref{eq:X_tilde_N_1_2} implies
    \begin{equation}\label{eq:X_tilde_X_bar_1}
        \begin{split}
            \mathbb{E}[|\overline{X}_{t} - \overline{Z}_{t}|^2]\leq& (1-C_3\Delta_l+C_2\Delta_l^2)^{(t-s)/\Delta_l}\mathbb{E}[|\overline{X}_{s} - \overline{Z}_{s}|^2]\\ +&(C_4\Delta_l+C_2\Delta_l^2)\sum_{k=0}^{(t-s)/\Delta_l}(1-C_3\Delta_l+C_2\Delta_l^2)^k\mathbb{E}[\mathcal{M}(\mu^{l,N}_{t-k\Delta_l},\mu^l_{t-k\Delta_l})^2]\\
            \leq& Ce^{(-C_3+C_2\Delta_l)(t-s)} + \zeta (C_4\Delta_l+C_2\Delta_l^2)\sum_{k=0}^{\infty}(1-C_3\Delta_l+C_2\Delta_l^2)^k\\
            \leq&C\zeta,
        \end{split}
    \end{equation}
    with $C$ independent of $\Delta_l$. Therefore 
    \begin{equation}\label{eq:X_tilde_X_bar_limit_1}
        \lim_{\substack{N\to\infty\\t\to\infty}}\mathbb{E}[|\overline{X}_{t} - \overline{Z}_{t}|^2]=0.
    \end{equation}
    Finally, we have that 
    $$
\mathbb{E}[\mathcal{W}_2(\Pi^l,\pi^l)^2]\leq 3\mathbb{E}[\mathcal{W}_2(\Pi^l,\mathcal{L}_{\overline{X}_t})^2] + 3\mathbb{E}[\mathcal{W}_2(\mathcal{L}_{\overline{X}_t},\mu^l_t)^2] + 3\mathcal{W}_2(\mu^l_t,\pi^l)^2.
$$
    The first term approaches $0$ as $t,N\to\infty$ by \eqref{eq:X_tilde_N_inq_1}, the second term approaches $0$ using \eqref{eq:X_tilde_X_bar_limit_1}, and the third term approaches $0$ by Theorem \ref{thm:ergod_1}.
\end{proof}

Finally, using Theorem \ref{thm:2nd} we have the following corollary when we allow $N=N_l$ (recall that $N_l$ is defined in the main text) to grow with $l$.
\begin{cor}\label{cor:main_cor}
    Assume \hyperref[assump:A1]{(A1-2)}. Then we have that
    $$
\mathcal{W}_2(\Pi^l,\pi)\xrightarrow[l\to\infty]{\mathbb{L}_2}0.
$$
\end{cor}
\begin{proof}
    By the triangular inequality we have
    $$
\mathcal{W}_2(\Pi^l,\pi)
\leq \mathcal{W}_2(\Pi^l,\pi^l) + \mathcal{W}_2(\pi^l,\pi).
$$
    By using \eqref{eq:X_tilde_X_bar_1} we have $\mathcal{W}_2(\Pi^l,\pi^l)\xrightarrow[l\to\infty]{\mathbb{L}_2}0$, and using Theorem \ref{eq:invaraint_delta_1} we have $
\mathcal{W}_2(\pi^l,\pi)\xrightarrow[l\to\infty]{}0$.
\end{proof}


\begin{thebibliography}{99}

\bibitem{siddig}
{\sc Awadelkarim}, E., \& {\sc Jasra}, A.~(2024).
Multilevel particle filters for partially observed McKean-Vlasov stochastic differential equations.
arXiv preprint.

\bibitem{ub_grad_new}
{\sc Awadelkarim}, E., {\sc Jasra}, A. \& {\sc Ruzayqat}, H.~(2024).
Unbiased parameter estimation for partially observed diffusions. 
\emph{SIAM J.  Control Optim. }, {\bf 62},  2664-2694.


\bibitem{baladron}
{\sc Baladron}, J., {\sc Fasoli}, D., {\sc Faugeras}, O., \& {\sc Touboul}, J..~(2012).
Mean-field description and propagation of chaos in networks of Hodgkin-Huxley and FitzHugh-Nagumo neurons. 
\newblock{\em The Journal of Mathematical Neuroscience}, \textbf{2}(10).




\bibitem{ml_is_mv}
{\sc Ben Rached,} N. , {\sc Haji-Ali}, A.-L., {\sc Shyam,} M., \& {\sc Tempone}, R. (2022).
Multilevel Importance Sampling for McKean-Vlasov Stochastic Differential Equation. arXiv preprint.
\textcolor{black}{
\bibitem{mi_is_mv}
{\sc Ben Rached,} N. , {\sc Haji-Ali}, A.-L., {\sc Shyam,} M., \& {\sc Tempone}, R. (2023).
Multi-Index Importance Sampling for McKean-Vlasov Stochastic Differential Equation.  arXiv preprint.
}
\textcolor{black}{
\bibitem{Bou}
{\sc Bou-Rabee, \& Schuh, K.}~(2023). 
Convergence of unadjusted Hamiltonian Monte Carlo for mean-field models. 
{\it Electronic Journal of Probability.}, \textbf{28}, 1--40.}

\bibitem{bucy}
{\sc Bucy}, R. S.~(1965) 
Nonlinear filtering theory. 
\newblock{\em IEEE Trans. Automat. Control}, \textbf{10}, 198.


%
%

\bibitem{chada1}
{\sc Chada}, N. K., {\sc Franks}, J., {\sc Jasra}, A., {\sc Law}, K. J. H. \& {\sc Vihola}, M. ~(2021). 
Unbiased estimation of discretely observed hidden Markov models. 
\emph{SIAM/ASA Journal on Uncertainty Quantification} \textbf{9}(2), 763--787.


\bibitem{chada}
{\sc Chada}, N. K., {\sc Leimkuhler}, B.,  {\sc Paulin}, D. \& {\sc Whalley},  P. A.~(2023).
Unbiased Kinetic Langevin Monte Carlo with Inexact Gradients.  arXiv preprint, arxiv:2311.05025.

\bibitem{pages}
{\sc Chassagneux},  J. F., \& {\sc Pages}, G.~(2024).
Computing the invariant distribution of McKean-Vlasov SDEs by ergodic simulation. arXiv preprint.
\textcolor{black}{
\bibitem{chen}
{\sc Chen}, X. \& {\sc Dos Reis},  G.~(2024).
 Euler simulation of interacting particle systems
and McKean-Vlasov SDEs with fully super-linear growth drifts in space and
interaction. 
\emph{\em IMA Journal of Numerical tAnalysis} 44(2), 751-796.}


\bibitem{cuia}
{\sc Cuia}, Y. , {\sc Lib}, X. \& {\sc Liu},  Y.~(2024).
Explicit numerical approximations for McKean-Vlasov stochastic differential equations in finite and infinite time.
arxiv preprint.
\textcolor{black}{
\bibitem{delmoral}
 {\sc Del Moral} P.~(2013).
\emph{Mean field simulation for Monte Carlo integration},
 Monographs on Statistics and Applied Probability, CRC Press, Boca Raton, FL, 2013.
978-1-4665-0405-9,3060209.}

\bibitem{du}
{\sc Du}, K., {\sc Jiang}, Y. \& {\sc Li}, J.~(2023).
Empirical approximation to invariant measures for McKean--Vlasov processes: Mean-field interaction vs self-interaction.
\emph{Bernoulli},  {\bf 29}(3), 2492--2518.


\bibitem{erban}
{\sc Erban}, R., {\sc Haskovec}, J. \& {\sc Sun}, Y.~(2016).
A Cucker--Smale model with noise and delay.
\newblock{\em SIAM J. App. Math.}, \textbf{76}, 151--184.


\bibitem{garnier}
{\sc Garnier}, G., {\sc Papanicolaou}, G. \& {\sc Yang}, W.~(2013).
Large deviations for a mean field model of systemic risk. 
\newblock{\em SIAM J. Financial Math.}, \textbf{4}, 151--184.


%
%
%
  


\bibitem{glynn2}
{\sc Glynn}, P.~W. \& 
{\sc Rhee}, C.~H.~(2014). 
Exact estimation for Markov chain equilibrium expectations. 
\emph{J. Appl. Probab.}, {\bf 51}, 
377--389.


\bibitem{ub_grad}
{\sc Heng}, J., {\sc Houssineau}, J. \& {\sc Jasra}, A.~(2024). On unbiased score estimation for partially observed diffusions.  \emph{J. Mach. Learn. Res.},  {\bf 25}, 1-66.


\bibitem{hk}
{\sc Hegselmann}, R. \&  {\sc Krause}, U.~(2014). 
Opinion dynamics and bounded confidence: models, analysis and simulation.
 \newblock{\em Journal of Artificial Societies and Social Simulation}, \textbf{5}(3), 2002.
 \textcolor{black}{
  \bibitem{HL23}
{\sc Hoffmann}, M. \&  {\sc Liu}, Y.~(2023). 
 A statistical approach for simulating the density solution of a McKean-Vlasov equation.
Arxiv preprint, arxiv:2305.06876, 2023.}
 
 
 




\bibitem{ub_pf}
{\sc Jasra}, A., {\sc Law}, K. J. H. \& {\sc Yu}, F.~(2022). Unbiased filtering of a class of partially observed diffusions.
\emph{Adv. Appl. Probab.} {\bf 54}, 661-687.
\textcolor{black}{
\bibitem{maama}
{\sc Jasra}, A., {\sc Maama}, M. \& {\sc Tempone}, R.~(2024).
Parameter estimation for partially observed McKean-Vlasov diffusions.
Arxiv preprint.}


\bibitem{kuntz}
{\sc Kuntz}, J.,  {\sc Lim} J. N. \&  {\sc Johansen}, A.~(2023). Particle algorithms for maxi- mum likelihood training of latent variable models. \emph{AISTATS 2023},  {\bf 206},  5134--5180.







\bibitem{lu}
{\sc Lu}, F., {\sc Maggioni}, M. \& {\sc Tang}, S.~(2022). 
Learning interaction kernels in stochastic systems of interacting particles from multiple trajectories.
\newblock{\em Found. Comput. Math.}, \textbf{22}, 1013--1067.


\bibitem{lu2}
{\sc Lu}, F., {\sc Maggioni}, M. \& {\sc Tang}, S.~(2021). 
Learning interaction kernels in heterogeneous systems of agents from multiple trajectories.
\newblock{\em J. Machine Learning Research}, \textbf{22}, 32 1--67.



\bibitem{mish}
{\sc Mishura}, Y \& {\sc Veretennikov}, A. Y.~(2021). 
Existence and uniqueness theorems for solutions of McKean--Vlasov stochastic equations.
\newblock{\em  Theory of Probability and Mathematical Statistics}, \textbf{103}, 59--101.





\bibitem{mck}
{\sc McKean}, H.P.~(1966). 
 A class of Markov processes associated with nonlinear parabolic equations. 
 \emph{Proceedings of the National Academy of Sciences of the United States of America}, {\bf 56}(6), 1907.


\bibitem{mcl}
{\sc McLeish}, D.~(2011). A general method for debiasing a Monte Carlo estimator. \emph{Monte Carlo Meth. Appl.}, {\bf 17}, 301--315.





\bibitem{motsch}
{\sc Motsch}, S. \& {\sc Tadmor}, E.~(2014). 
 Heterophilious Dynamics Enhances Consensus. 
 \emph{SIAM Review}, \textbf{56}(4), 577--621.
\textcolor{black}{
\bibitem{reis}
G {\sc dos Reis}, G., {\sc Smith}, G. \& {\sc Tankov}, P.~(2023).
Importance sampling for McKean-Vlasov SDEs..
\emph{Applied Mathematics and Computation}, \textbf{453}, 128078.
}

\bibitem{chada2}
 {\sc Paulin}, D., {\sc Whalley} P. A., {\sc Chada}, N. K.  \& {\sc Leimkuhler}, B.~(2025).
Sampling from Bayesian neural network posteriors with symmetric minibatch splitting Langevin dynamics.  
28th International Conference on Artifical Intelligence and Statistics, PMLR, \textbf{258}, 5014--5022.





\bibitem{rhee}
{\sc Rhee}, C. H. \& {\sc Glynn}, P.~(2015). Unbiased estimation with square root convergence for SDE models. \emph{Op. Res.},~{\bf 63}, 1026--1043. 


\bibitem{langevin}
{\sc Ruzaqyat}, H., {\sc Chada}, N. K. \& {\sc Jasra}, A.~(2023). Unbiased estimation using underdamped Langevin dynamics.  \emph{SIAM J. Sci. Comp.},  {\bf 45},  A3047-A3070.
\textcolor{black}{
\bibitem{sharrock}
{\sc Sharrock}, L., {\sc Kantas}, N.,  {\sc Parpas}, P. \& {\sc Pavliotis}, G. A.~(2023). Online parameter estimation for the McKean-Vlasov stochastic differential equation.  \emph{Stochastic Process. Appl.},  {\bf 162},  481--546.}



\bibitem{basic_method}
{\sc Sznitman},  A. S.~(1991). Topics in propagation of chaos. In \emph{Ecole d'Ete de Probabilites de Saint-Flour XIX}.  {\bf 1464}, 165--251.






\bibitem{wang}
{\sc Wang},  F. Y.~(2018)
Distribution dependent SDEs for Landau type equations.
\emph{Stochastic Processes and their Applications}.  {\bf 128}(2), 595--621.


\bibitem{yang}
{\sc Yang}, H. , {\sc Hasan}, A.,  {\sc Ng}, Y. \& {\sc Tarokh},  P. V.~(2024).
Neural McKean-Vlasov processes: distributional dependence in diffusion processes.
\newblock{\em AISTATS (to appear)}.




\end{thebibliography}
\end{document}